\documentclass[aps,prd,preprint,floats,epsf,superscriptaddress,nofootinbib,amsmath]{revtex4}
\usepackage{graphicx,color,rotating}
\usepackage{hyperref}
\usepackage{epsfig,color}
\usepackage{slashed}

\begin{document}

\bigskip

\title{Probing a dark photon using rare leptonic kaon and pion decays}

\author{Cheng-Wei Chiang}
\email[e-mail: ]{chengwei@phys.ntu.edu.tw}
\affiliation{Department of Physics, National Taiwan University, Taipei, Taiwan 10617, R.O.C.}
\affiliation{Institute of Physics, Academia Sinica, Taipei, Taiwan 11529, R.O.C.}
\affiliation{Physics Division, National Center for Theoretical Sciences, Hsinchu, Taiwan 30013, R.O.C.}

\author{Po-Yan Tseng}
\email[e-mail: ]{poyen.tseng@ipmu.jp}
\affiliation{Kavli IPMU (WPI), UTIAS, The University of Tokyo, Kashiwa, Chiba 277-8583, Japan}

\date{\today}

\begin{abstract}
Rare leptonic kaon and pion decays $K^+(\pi^+) \to \mu^+\, \nu_{\mu}\, e^+ e^-$ can be used to probe a dark photon of mass ${\cal O}(10)$~MeV, with the background coming from the mediation of a virtual photon.  This is most relevant for the 16.7-MeV dark photon proposed to explain a 6.8$\sigma$ anomaly recently observed in $^8$Be transitions by the Atomki Collaboration.  We evaluate the reach of future experiments for different scenarios of how the dark photon couples with the standard model particles, and show that a great portion of the preferred 16.7-MeV dark photon parameter space can be decisively probed.  We also show the use of angular distributions to further distinguish the signal from the background.
\end{abstract}

\pacs{}

\maketitle

\section{Introduction~\label{sec:intro}}


The search of new gauge interactions has been of great interest.  Such efforts help us understand whether there is any other new force in Nature and how it fits to the grand picture of particle physics.  If found, there is also a possibility that the new force carrier provides a portal between the standard model (SM) visible sector and a hidden sector involving new dynamics and matter contents.

In the simplest scenario, such a new vector boson may have an origin from some extra $U(1)'$ gauge symmetry, under which some particles in both visible and hidden sectors are charged.  It may even have kinetic or mass mixing with the SM photon or $Z$ boson field to facilitate the mediation.  After its symmetry breaking, as is often assumed, the $U(1)'$ gauge boson acquires a mass and is commonly called the $Z'$ boson if it is heavier than the electroweak scale or the dark photon if it is lighter.  There have been a vast amount of studies on the neutral gauge boson over the years~\cite{Leike:1998wr,Langacker:2008yv,Rizzo:2006nw,Erler:2009jh,Borodatchenkova:2005ct,Fayet:2007ua,Batell:2009yf,Essig:2009nc,Reece:2009un,Essig:2013lka,Alexander:2016aln}.  For example, recent direct searches at the LHC have already pushed the lower bound on the sequential $Z'$ mass to about 3~TeV~\cite{Aaboud:2016cth,Khachatryan:2016zqb}.  Further investigations rely on advances in the colliding energy and beam luminosity.

On the other hand, probes of the dark photon of mass below sub-GeV have been done using nuclear transitions~\cite{Treiman:1978ge,Savage:1986ty,Hallin:1986gh,Savage:1988rg} or from its effects on the magnetic dipole moment of electron and/or muon~\cite{Pospelov:2008zw,Davoudiasl:2012ig,Endo:2012hp}.  The dark photon can be radiated off from some particle by bremsstrahlung and then decay into a pair of leptons in the fixed-target and beam-dump experiments~\cite{Merkel:2011ze,Abrahamyan:2011gv,Bjorken:2009mm,Blumlein:2011mv,Gninenko:2011uv,Gninenko:2012eq,Andreas:2012mt,Blumlein:2013cua}.  With an appropriate coupling, it can be produced at $e^+e^-$ colliders as well.  If the dark photon couples to quarks, one can consider its production in meson decays, if kinematically allowed.

Recently, there is an elevated interest in the study of dark photon because of an experimental anomaly involved in isoscalar $^8$Be transitions reported by Krasznahorkay et.~al.~\cite{atomki}.  In the transition from an excited state to the ground state, the nucleus emitted an electron-positron pair whose open angle and invariant mass were found to deviate from the SM expectation of internal pair creation (IPC) by $6.8\sigma$.  It was shown that the distributions of open angle and invariant mass could be well fit by introducing a new particle with mass 16.7~MeV produced in the transition.  In Refs.~\cite{jonathan_1,jonathan_2}, the authors claimed that the new particle could be a vector boson $X$ and provided the preferred ranges of its couplings with SM particles that were consistent with current dark photon search constraints.
There are several analyses on how to further test the model in Ref.~\cite{jonathan_1} using low-energy physical processes~\cite{Chen:2016dhm,Chen:2016kxw}, as well as proposals of alternative models for the $^8$Be anomaly~\cite{Gu:2016ege,Kitahara:2016zyb,Ellwanger:2016wfe,Chen:2016tdz,Neves:2016nek,Kahn:2016vjr,Seto:2016pks}.

In this work, we propose to use rare leptonic decays of kaon and pion, $K^+ \to  \mu^+\, \nu_{\mu}\,  e^+ e^-$ and $\pi^+ \to  \mu^+\, \nu_{\mu}\,  e^+ e^-$, as felicitous means to probe the light dark photon in the mass range of about ${\cal O}(10)$~MeV, particularly in view of the putative gauge boson $X$ with vectorial couplings mentioned above.  This is because the final-state electron-positron pair production can be enhanced via the mediation of the $X$ boson over the SM background through a virtual photon.  We show that the SM background and the signal have very different spectra in the $e^+e^-$ invariant mass: the SM background has a continuous spectrum, whereas the decay through a dark photon features in a sharp resonance peak around the dark photon mass $m_X$.  Moreover, the $K^+ \to \mu^+\, \nu_{\mu}\, e^+ e^-$ and $\pi^+ \to \mu^+\, \nu_{\mu}\, e^+ e^-$ decays are able to probe most part of the preferred coupling space inferred from the $^8$Be anomaly, assuming specific production rates of kaons/pions and their energy resolutions.  We provide the projected experimental limits on the dark photon couplings based on these decay processes.  We show that the signal events and SM background events have different behaviours in the $\mu^+\nu$ angular distribution in the leptonic kaon decay.  We also discuss the influence of the kaon/pion structure dependence ({\i.e.}, the form factors), and find their effects almost irrelevant in current considerations.

This paper is organized as follows.  In Section~\ref{sec:X}, we review the leptonic meson decays of interest to us.  We present their decay amplitudes, including both inner bremsstrahlung and structure-dependent parts, and provide the form factors involved in the latter part.  In Section~\ref{sec:scen}, we consider three different dark photon scenarios, depending on how it couples to the SM particles, and three different experimental schemes for the kaon and pion decays.  We estimate the projected reach in the dark photon couplings in Section~\ref{sec:proj}.  Numerical results are presented in Section~\ref{sec:resu}, where we also discuss effects coming from the structure-dependent contributions.  Our findings are summarized  in Section~\ref{sec:summary}.

\section{The $K^+/\pi^+ \to \mu^+ \nu_{\mu} e^+ e^-$ decays via a dark photon and within the standard model~\label{sec:X}}

There can be many different ways to realize a light $U(1)'$ gauge boson and let it couple with SM particles (directly or via mixing with the photon and/or $Z$ boson).  In the following, we will keep the formalism as general as possible, without explicitly referring to any particular new physics model.  Suppose the gauge coupling $e'$ of the dark photon $X$ has a ratio of $\varepsilon$ to the SM electric coupling, {\it i.e.}, $e' = \varepsilon e$.  We denote the coupling charge of electron, muon, $u,d,s$ quarks, and $K^+,\pi^+$ mesons respectively by $Q_{e,\mu,u,d,s,K^+,\pi^+}$, whose values depend on the model.

The dark photon $X$ can contribute to the $K^+(k) \to \mu^+(\ell) \,\nu_{\mu}(q)\, X(q') \to \mu^+(\ell)\, \nu_{\mu}(q)\,  e^+(\ell_1) e^-(\ell_2)$ decay, where the variables in the parentheses denote the momenta of the corresponding particles.  The radiative kaon decay involves both inner bremsstrahlung (IB) and structure-dependent (SD) parts, and the total decay amplitude is given by~\cite{form_factor}
\begin{equation}
\label{eq1}
i {\cal M}_{K} =
\frac{G_F}{\sqrt{2}} V_{us}^*\,  (\varepsilon e Q_{K^+})\, \epsilon^*_{\rho}(q') 
\left[
f_K\, \bar{L}^{\rho}-\sqrt{2} \bar{H}^{\rho \mu}\, \ell_{\mu}
\right] ~,
\end{equation}
where $G_F$ is the Fermi decay constant, $f_K=155.6$~MeV is the kaon decay constant, the Cabibbo-Kobayashi-Maskawa (CKM) matrix element $V_{us}\simeq 0.22538$, and
\begin{eqnarray}
\label{eq2}
\bar{L}^{\rho} &=& m_{\mu}
\bar{u}(q)(1+\gamma_5)\left\lbrace 
\frac{2k^{\rho}-q'^{\rho}}{2k\cdot q'-q'^2}+
\left(\frac{Q_{\mu}}{Q_{K^+}}\right) 
\frac{2\ell^{\rho}+\slashed{q}'\gamma^{\rho}}{2\ell\cdot q'+{q'}^2}
\right\rbrace v(\ell) ~,
\nonumber \\
\ell^{\mu} &=& \bar{u}(q)\gamma^{\mu}(1-\gamma_5) v(\ell) ~,
\nonumber \\
\bar{H}^{\rho \mu}&=& i V_1 \epsilon^{\rho \mu \alpha \beta} q'_{\alpha}k_{\beta}
                  -A_1 \left( q'\cdot W g^{\rho \mu}-W^{\rho}q'^{\mu} \right)
                  \nonumber \\
               && -A_2 \left( q'^2 g^{\rho \mu}-q'^{\rho}q'^{\mu} \right)
                  -A_4 \left(q'\cdot W q'^{\rho}-q'^2W^{\rho} \right)W^{\mu} ~,
\nonumber \\
\epsilon^*_{\rho}(q')
&=&
\frac{Q_e\, \varepsilon e}{q'^2-m^2_{X}+im_{X}\Gamma_{X}}
\left[ \bar{u}(\ell_2)\gamma_{\rho}v(\ell_1) \right]\,,
\end{eqnarray}
with the $\bar{L}^{\rho}$ part being due to IB, the $\bar{H}^{\rho \mu}$ part containing the SD form factors, $W^{\mu}\equiv k^{\mu}-q'^{\mu}$, $q'\equiv\ell_1+\ell_2$, and $\Gamma_X$ denoting the total width of $X$.  The expressions in Eqs.~(\ref{eq1}) and (\ref{eq2}) can be readily modified to give those for the SM background by replacing symbols associated with the dark photon by those for photon.

We follow the convention in Ref.~\cite{form_factor} for the kaon form factors, which is consistent with the one used in Ref.~\cite{pdg}.
\footnote{
We have compared three references~\cite{form_factor,pdg,form_factor_2002} and decided to mainly follow the convention in Ref.~\cite{form_factor}.  It is consistent with Ref.~\cite{pdg} after taking into account an overall factor of $-\sqrt{2}m_K$, as explicitly shown in Eq.~(C.8) of Ref.~\cite{form_factor}.
}
We set $A_4=0$, and the rest form factors from recent measurements are parametrized as~\cite{form_factor_2002,kaon1}
\begin{eqnarray}
\sqrt{2}m_K\,A_1(q'^2,W^2)&=& \frac{-F_A}{\left(1-q'^2/m_{\rho} \right)\left(1-W^2/m_{K_1} \right)}\,,  \nonumber \\
\sqrt{2}m_K\,A_2(q'^2,W^2)&=& \frac{-R}{\left(1-q'^2/m_{\rho} \right)\left(1-W^2/m_{K_1} \right)}\,,  \nonumber \\
\sqrt{2}m_K\,V_1(q'^2,W^2)&=& \frac{-F_V}{\left(1-q'^2/m_{\rho} \right)\left(1-W^2/m_{K^*} \right)}\,,
\end{eqnarray}
with $F_A=0.031$, $R=0.235$, $F_V=0.124$, $m_{\rho}=770$~MeV, $m_{K_1}=1270$~MeV, and $m_{K^*}=892$~MeV.

In the case of signal events, the distribution in the electron-positron invariant mass is a sharp resonance peak around $m_X$.  The partial width of the dark photon decay into a pair of fermions is
\begin{equation}
\Gamma_{X \to f\bar{f}}=\frac{N_C m_{X}}{12 \pi}
\sqrt{1-4\,r_f}\times
\left[ g^2_V(1+2\,r_f)+g^2_A(1-4\,r_f) \right]\,,
\end{equation}
where $r_f=m^2_f/m^2_{X}$, $N_C$ is the color factor, and $g_V$, $g_A$ are the vector and axial-vector couplings between $X$ and $f$.  Taking $m_X=16.7$~MeV and assuming vectorial couplings as in Refs.~\cite{jonathan_1,jonathan_2}, the dominant decay channel should be $e^+e^-$, $g_V=e\cdot\varepsilon\cdot Q_e$ and $g_A=0$.
For the SM background, one should set $\varepsilon = 1$, $Q_{K^+}=1$, $Q_{\mu}=-1$ and replace $\epsilon^*(q')$ by
\begin{equation}
\epsilon^*_{\rho}(q')=\frac{e}{q'^2}
\left[ \bar{u}(\ell_2)\gamma_{\rho}v(\ell_1) \right]\,.
\end{equation}
In this case, the electron-positron invariant mass distribution gives a continuous spectrum.

We also consider the pion decay $\pi^+(k) \to \mu^+(\ell)\,\nu_{\mu}(q)\, X(q') \to \mu^+(\ell)\, \nu_{\mu}(q)\, e^+(\ell_1) e^-(\ell_2)$.  Similar to the kaon decay, the pion decay amplitude is~\cite{bryman,pion_ff}
\begin{equation}
\label{eq7}
i {\cal M}_{\pi} =
\frac{G_F}{\sqrt{2}} V_{ud}\,  (\varepsilon e Q_{\pi^+})\, \epsilon^*_{\rho}(q') 
\left\lbrace
f_{\pi}\, \bar{L}^{\rho}-\bar{H}^{\rho \mu}\, \ell_{\mu}
\right\rbrace
\end{equation}
where
\begin{eqnarray}
\label{eq8}
\bar{L}^{\rho} &=& m_{\mu}
\bar{u}(\ell)(1-\gamma_5)\left\lbrace 
\frac{2k^{\rho}-q'^{\rho}}{2k\cdot q'-q'^2}+
\left(\frac{Q_{\mu}}{Q_{{\pi}^+}}\right) 
\frac{2\ell^{\rho}+\slashed{q}'\gamma^{\rho}}{2\ell\cdot q'+{q'}^2}
\right\rbrace v(q) ~,
\nonumber \\
\ell^{\mu} &=& \bar{u}(\ell)\gamma^{\mu}(1-\gamma_5) v(q) ~,
\nonumber \\
\bar{H}^{\rho \mu}&=& i \frac{F_V}{m_{\pi}} \epsilon^{\rho \mu \alpha \beta} q'_{\alpha}k_{\beta}
-\frac{F_A}{m_{\pi}} \left( q'\cdot k\, g^{\rho \mu}-k^{\rho}q'^{\mu} \right) ~,
\end{eqnarray}
where $m_{\pi}=139.57$~MeV, $f_{\pi}=130$~MeV is the pion decay constant, the CKM matrix element $V_{ud}\approx0.974$, and the form factors~\cite{pion}
\begin{eqnarray}
F_V(\bar{q}^2) &=& F_V(0)\times (1+a\bar{q}^2)\,
\nonumber \\
F_A(\bar{q}^2) &=& F_A(0)\,
\end{eqnarray}
with $F_V(0)=0.0258$, $F_A(0)=-0.0117$, $a=0.10$, and $\bar{q}^2=1-(2E_{\gamma'}/m_{\pi})$.

The IB term in the $\pi^+ \to \mu^+ \nu e^+e^-$ decay amplitude is proportional to $m_{\mu}$, therefore the IB contribution is more important than that in the $\pi^+ \to e^+ \nu e^+e^-$ decay amplitude.  The SD contribution of the $\pi^+ \to e^+ \nu e^+e^-$ becomes more important at the kinematic regime of large $e^+e^-$ invariant mass~\cite{pion}.

\section{Dark photon scenarios and search schemes~\label{sec:scen}}

To explain the $^8{\rm Be}$ anomaly, it is sufficient for the dark photon to have couplings with only fermions in the first family while satisfying gauge anomaly cancellation~\cite{atomki,jonathan_1,jonathan_2}.  The preferred ranges of dark photon couplings are found to be~\cite{jonathan_1}
\begin{eqnarray}
\label{eq23}
|\varepsilon Q_n| &=& (2-10)\times 10^{-3}\,, \nonumber \\
|\varepsilon Q_p| &\lesssim& 1.2\times 10^{-3}\,, \nonumber \\
|\varepsilon Q_e| &=& (0.2-1.4)\times 10^{-3}\,, \nonumber
\end{eqnarray}
where the coupling preferably protophobic in order to evade the constraints from the $\pi^0 \to X \gamma$ decays measured at the NA48/2 experiment.

If the dark photon also couples to the second family and we require gauge anomaly cancellation within each family, then the $U(1)'$ charges for muon and strange quark have either $Q_{\mu}=+Q_e,\,Q_s=+Q_d$ or $Q_{\mu}=-Q_e,\,Q_s=-Q_d$.  We will thus consider the following three dark photon scenarios:
\begin{itemize}

\item {\bf Scenario 1}: Only $Q_{u,d,e}\neq 0$.

\item {\bf Scenario 2}: Add $Q_{\mu}=+Q_e$ and $Q_s=+Q_d$ to Scenario 1.

\item {\bf Scenario 3}: Add $Q_{\mu}=-Q_e$ and $Q_s=-Q_d$ to Scenario 1.

\end{itemize}
With $Q_u=(\frac23 Q_p - \frac13 Q_n)$, $Q_d=(\frac23 Q_n - \frac13 Q_p)$, $Q_{K^+} = Q_u-Q_s$, and $Q_{\pi^+} = Q_u-Q_d$, the $U(1)'$ charge of kaon is $Q_{K^+} = \frac{2}{3}Q_p - \frac{1}{3}Q_n$ for Scenario~1, $Q_p - Q_n$ for Scenario~2, and $\frac{1}{3}Q_p + \frac{1}{3}Q_n$ for Scenario~3.  The $U(1)'$ charge of pion is $Q_{\pi^+} = Q_p - Q_n$ for all three scenarios.

Here we consider three different experimental schemes for the estimation of projected limits in the determination of dark photon gauge coupling.  Suppose we produce $N_{K^+,\pi^+}$ kaons and pions, respectively, and measure the $e^+e^-$ invariant mass from the decays with an energy resolution of $\delta m$.  The schemes are~\footnote{Such kaon production and mass resolution may be achieved by the rare kaon decay experiment at JPARC.  The numbers for pions are based upon the stopped pion experiment at PIBETA during 1999-2001 and in 2004~\cite{pion}.}:
\begin{itemize}
\item {\bf Scheme 1}: $N_{K^+}=10^{12}$, $N_{\pi^+}=10^{14}$, and $\delta m=1$ MeV,

\item {\bf Scheme 2}: $N_{K^+}=10^{11}$, $N_{\pi^+}=10^{13}$, and $\delta m=1$ MeV,

\item {\bf Scheme 3}: $N_{K^+}=10^{11}$, $N_{\pi^+}=10^{13}$, and $\delta m=5$ MeV.
\end{itemize}
The energy resolution $\delta m$ used here is much larger than the dark photon width for $|e\, \varepsilon\, Q_e|\approx (0.2-1.4)\times 10^{-3}$, as preferred by the $^8{\rm Be}$ anomaly~\cite{jonathan_1,jonathan_2}.  Therefore, we cannot determine its width from the resonance peak in the $e^+e^-$ invariant mass distribution.  But in contrast, SM background gives a continuous spectrum.

\section{The projected limits~\label{sec:proj}}

%
\begin{table}[t!]
  \caption{\small \label{t1}
Branching ratios of the SM background in the range of $m_X - \frac{\delta m}{2}< m_{e^+e^-} < m_X + \frac{\delta m}{2}$, where $m_X=16.7$~MeV, for different experimental schemes.
}
\smallskip
\begin{ruledtabular}
\begin{tabular}{l c c c}
   & Scheme 1 & Scheme 2 & Scheme 3 \\
\hline
$\Delta {\rm BR}_{\gamma^*}(K^+ \to \mu^+ \nu e^+e^-)$ &  $2.54\times 10^{-7}$   & $1.29\times 10^{-6}$ & $2.70\times 10^{-6}$ \\
$\Delta {\rm BR}_{\gamma^*}(\pi^+ \to \mu^+ \nu e^+e^-)$ &  $1.61\times 10^{-10}$   & $8.69\times 10^{-10}$ & $2.21\times 10^{-9}$ \\
\end{tabular}
\end{ruledtabular}
\end{table}

In this section, we estimate the projected limits for the $U(1)'$ gauge coupling or, equivalently, the $\varepsilon$ parameter for the above-mentioned three experimental schemes, following the steps outlined in Ref.~\cite{Essig:2010xa}.

First, we use the simple definition of standard deviation
\begin{equation}
\label{eq10}
\sigma = \frac{S}{\sqrt{B}} ~,
\end{equation}
where $S$ is the number of signal events, and $B$ is that number of background events.  In the following, we will consider $\sigma = 2$, corresponding to about $95\%$ confidence limit (C.L.).  Base on the experimental schemes, we write down the ratio of $S$ and $B$ in terms of the branching ratios
\begin{equation}
\label{eq11}
\frac{S}{B}=\frac{{\rm BR}_{X}(\varepsilon,m_X)\times N_{K^+}}{\Delta{\rm BR}_{\gamma^*}\times N_{K^+}}\,,
\end{equation}
where ${\rm BR}_{X}(\varepsilon,m_X)$ and $\Delta{\rm BR}_{\gamma^*}$ are the branching ratios of the leptonic kaon (or pion) decay through $X$ and $\gamma^*$, respectively, with the requirement that the $e^+e^-$ invariant mass falls within the energy range: $m_X - \frac{\delta m}{2}< m_{e^+e^-} < m_X + \frac{\delta m}{2}$.  With a fixed $m_X$, ${\rm BR}_{X}(\varepsilon,m_X)$ is proportional to $\varepsilon^2$ due to the Breit-Wigner description and $[(q'^2-m^2_{X})^2+m^2_{X}\Gamma^2_{X}]^{-1} \simeq \frac{\pi}{m_{X}\Gamma_{X}}\delta(q'^2-m^2_{X})$.  Combining Eqs.~(\ref{eq10}) and (\ref{eq11}), we obtain the 2$\sigma$ probing limit on the ratio
\begin{equation}
\varepsilon^2_{\rm limit}
=\frac{2}{\sqrt{\Delta{\rm BR_{\gamma^*}}\times N_{K^+}}}\times
\frac{\tilde{\varepsilon}^2 \times \Delta{\rm BR}_{\gamma^*}}{{\rm BR}_{X}(\tilde{\varepsilon})}\,,
\end{equation}
where $\tilde{\varepsilon}$ can be any reasonable value, as long as $\Gamma_X \ll \delta m$.  The branching ratios of SM background in various schemes are listed in Table~\ref{t1}.

\begin{figure}[t!]
\includegraphics[height=2.2in,angle=0]{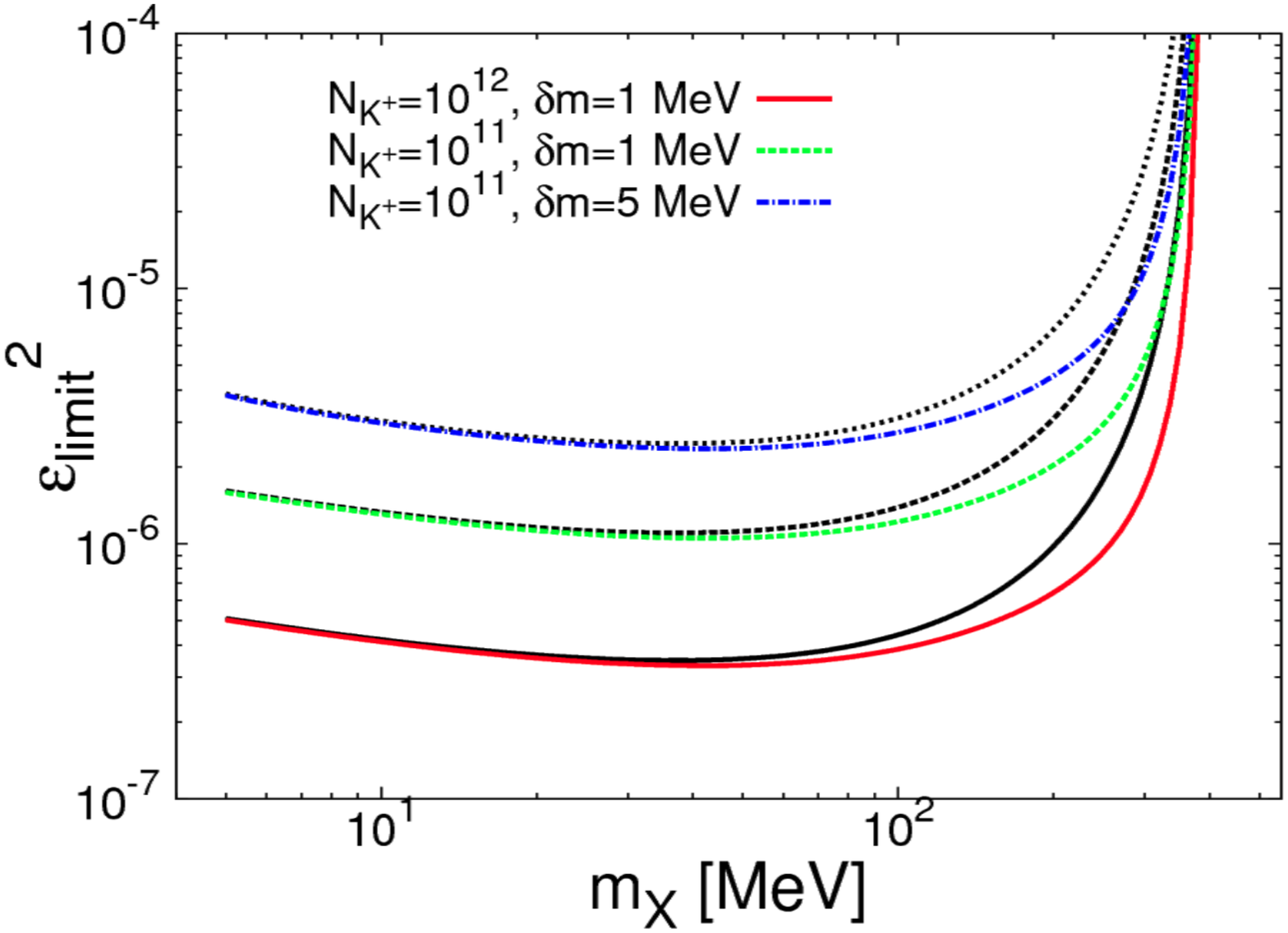}
\includegraphics[height=2.2in,angle=0]{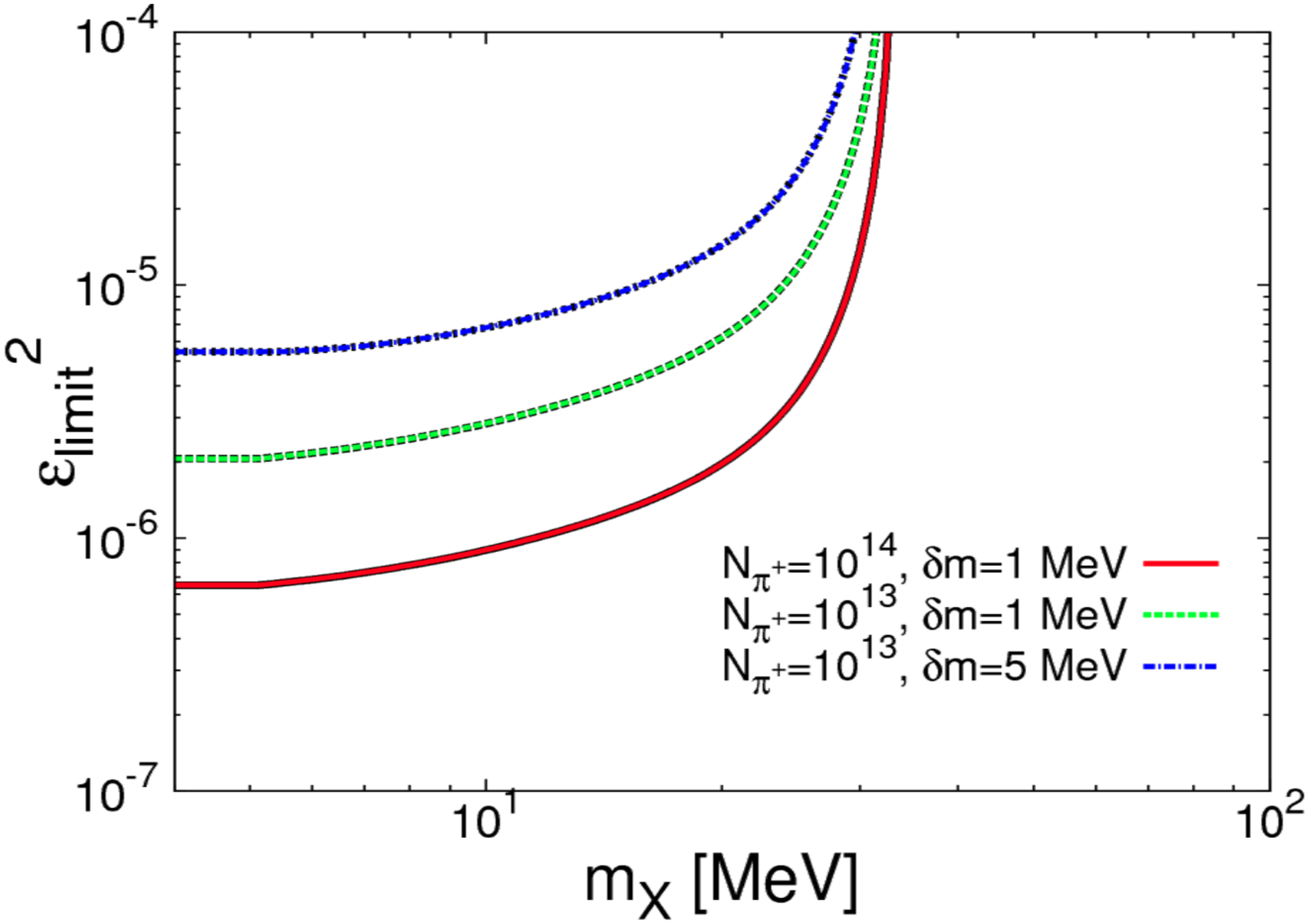}
\caption{\small \label{f1} 
Projected limits on $\varepsilon$ with $S/\sqrt{B}=2$ (about 95\% C.L.) with $Q_{K^+,\pi^+}=1$, and $Q_{e,\mu}=-1$ (Scenario 2). The left panel is for the $K^+$ decay, and the right panel is for the $\pi^+$ decay.  The red, green, and blue curves, including both IB and SD contributions, are for Schemes 1, 2, and 3, respectively.  For each colored curve, the adjacent black curves include only the IB contribution.
}
\end{figure}

In Fig.~\ref{f1}, we show $\varepsilon^2_{\rm limit}$ versus $m_X$ in colored curves for the three experimental schemes from kaon (left plot) and pion (right plot) decays by fixing $Q_{K^+,\pi^+}=1$ and $Q_{e,\mu}=-1$ ({\it i.e.}, Scenario 2).  We also show in black curves the effect of form factors on $\varepsilon^2_{\rm limit}$ by removing the SD part.  For the dark photon mass of interest to the $^8$Be anomaly, $m_X \sim 16.7$~MeV, the effects from kaon or pion form factors are basically negligible.

Before concluding this section, we comment on the decay length of the dark photon as produced from the kaon or pion decay.  The decay length of $X$ produced from $K^+$ is $\gamma_{X} \tau_{X} c$, where $\gamma_{X}$ is the boost factor, $\tau_{X}=1/\Gamma_{X \to ee}$ is the lifetime, and $c$ is the speed of light.  The largest boost factor from $K^+ \to \mu^+ \nu X$ is $\gamma_{X}\sim m_{K^+}/(2m_{X}) \sim 14.8$.  Taking the smallest value of $\varepsilon Q_e=0.2\times 10^{-3}$ for the $^8$Be anomaly, the largest decay length of $X$ with $m_X=16.7$ MeV is 
$$
\gamma_{X} \tau_{X} c \simeq 0.179\,\, {\rm cm}\,.
$$
Therefore, the dark photon should decay within the detector.  On the other hand, a decay length of $X$ of $\sim$ 1~cm corresponds to $\varepsilon Q_e\sim 8.5\times 10^{-5}$.

\section{Numerical results~\label{sec:resu}}

\begin{figure}[h]
\includegraphics[height=3.5in,angle=0]{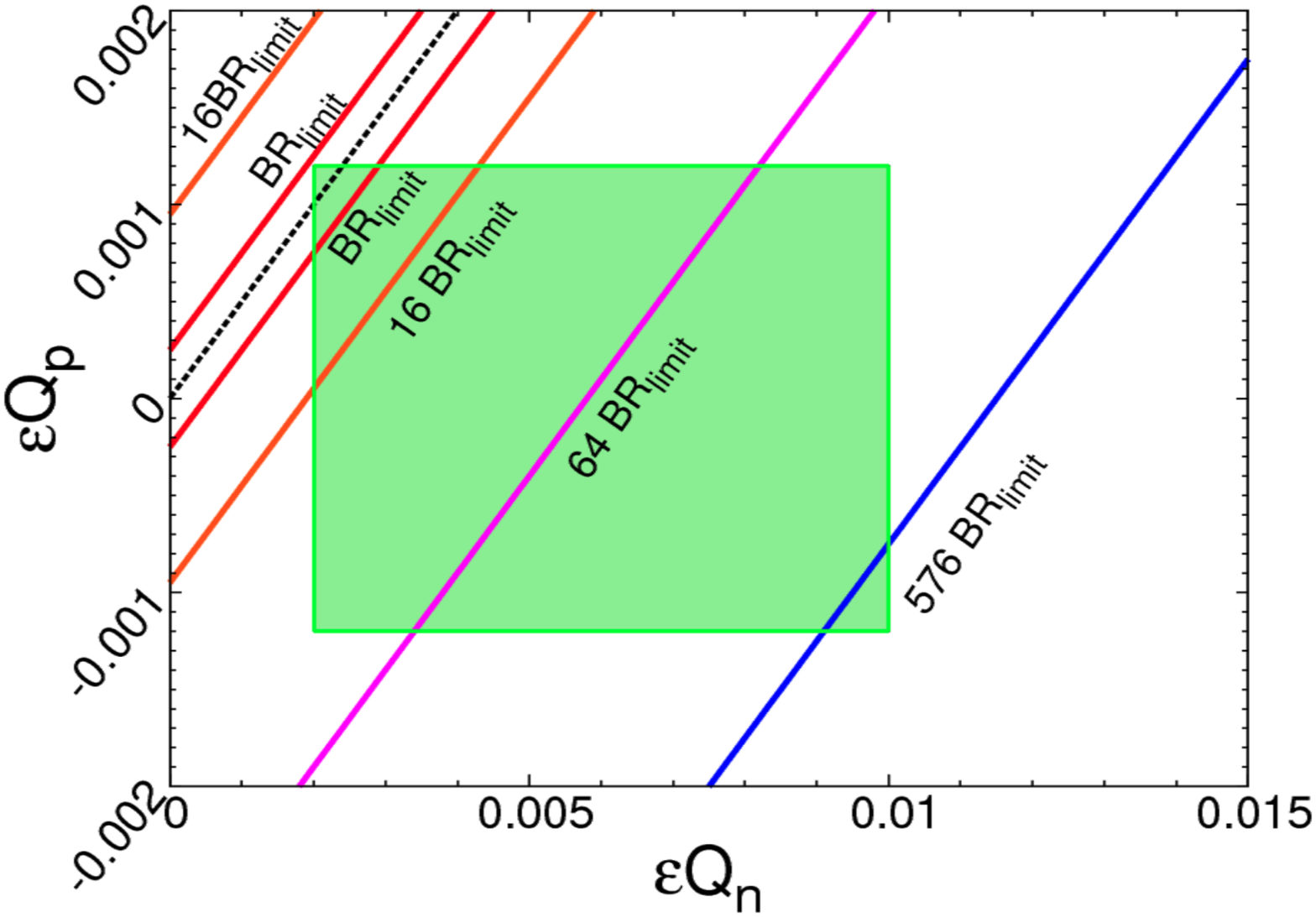}
\caption{\small \label{f2} 
Contours of ${\rm BR}_{X}$ for Scenario~1 on the $\varepsilon Q_n$-$\varepsilon Q_p$ plane.  ${\rm BR}_{\rm limit}$ comes from the experimental Scheme~1.  The light green area is the region favored by the $^8$Be anomaly~\cite{jonathan_1,jonathan_2}.  The black dotted line is for $\varepsilon Q_{K^+}=0$.  The colored lines indicate the contours for various values of ${\rm BR}_{X}$; for example, the red contour is for ${\rm BR}_{X}={\rm BR}_{\rm limit}=1.009\times 10^{-9}$.
}
\end{figure}
\begin{figure}[h]
\includegraphics[height=1.5in,angle=0]{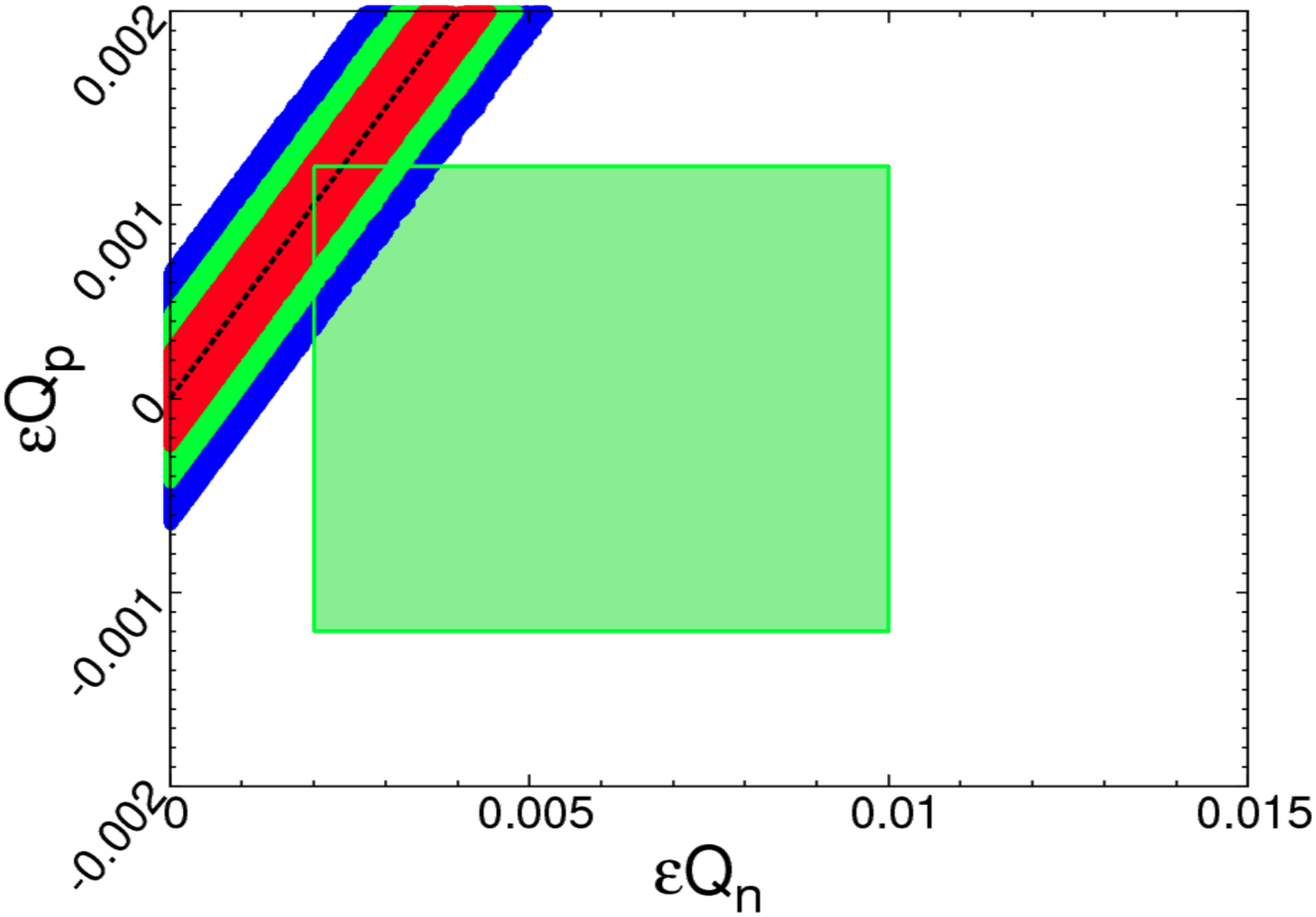}
\includegraphics[height=1.5in,angle=0]{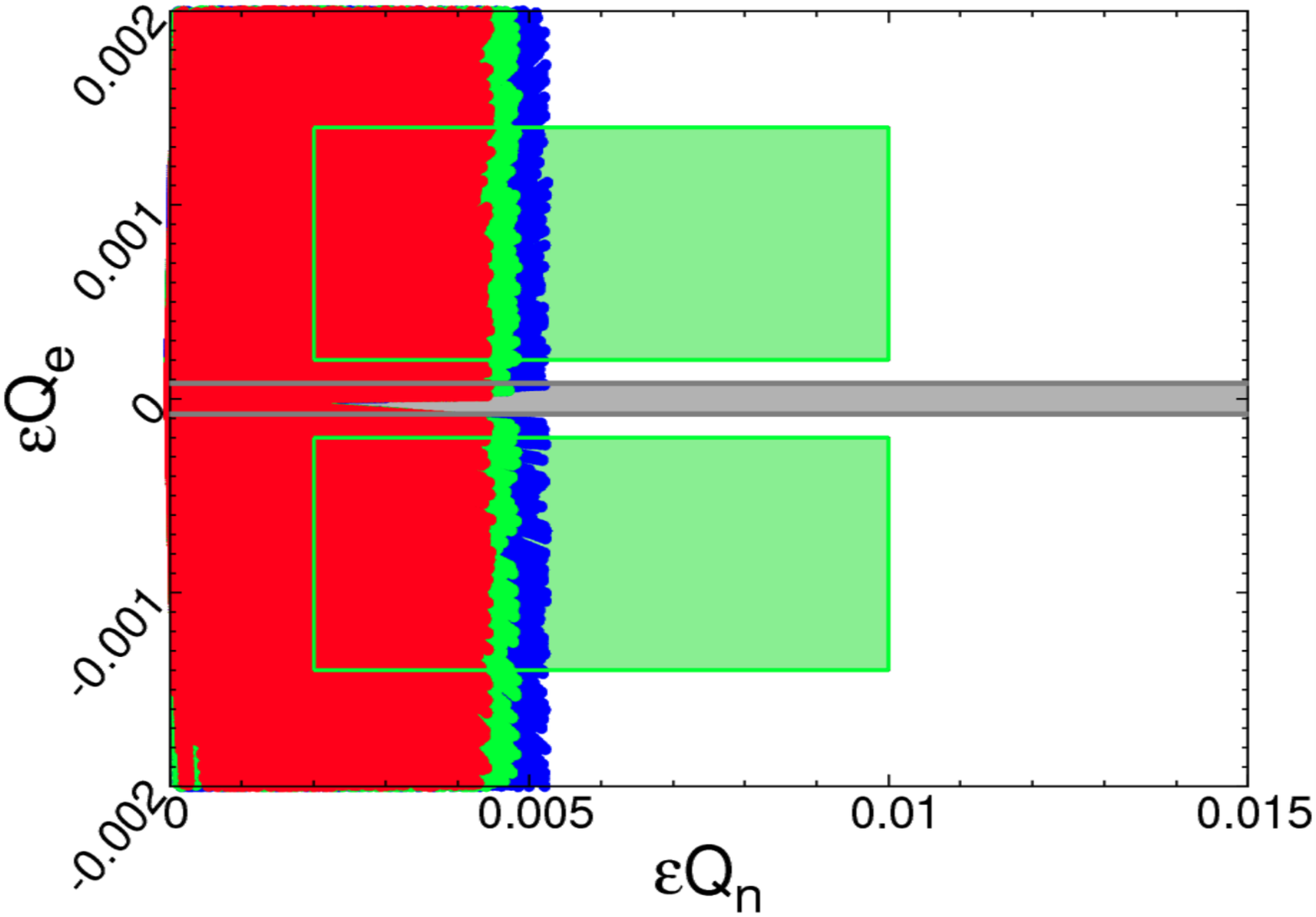}
\includegraphics[height=1.5in,angle=0]{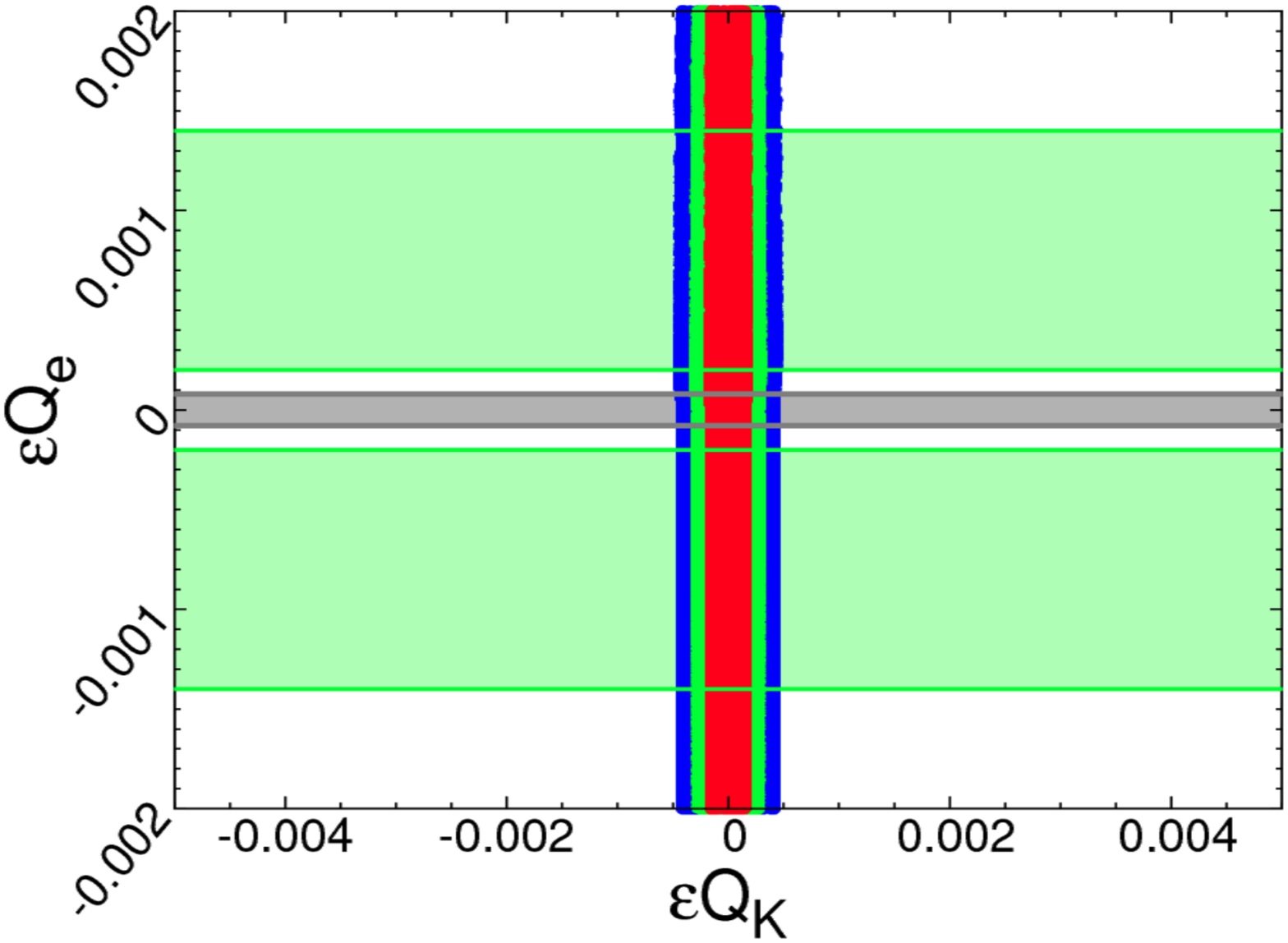}
\includegraphics[height=1.5in,angle=0]{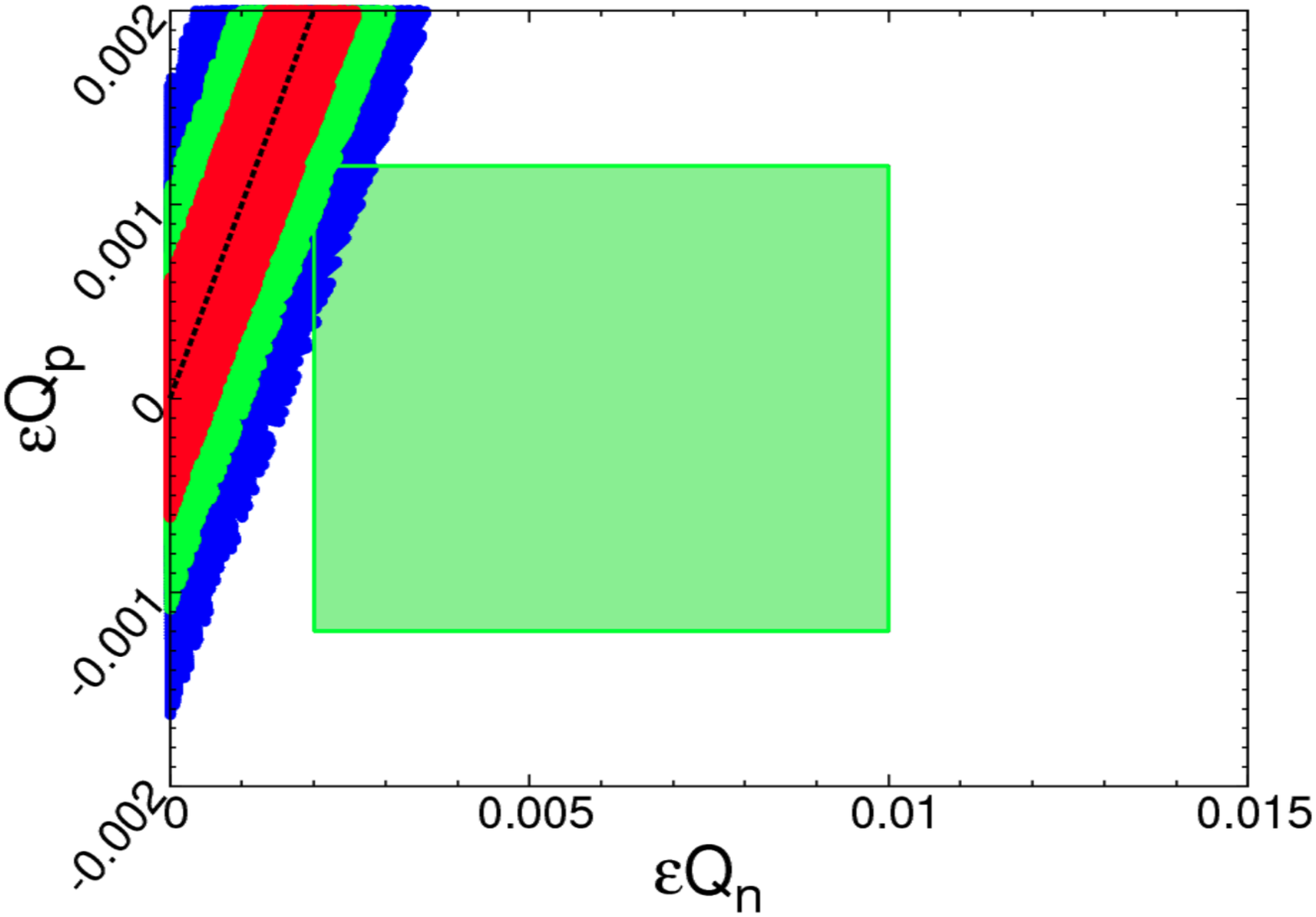}
\includegraphics[height=1.5in,angle=0]{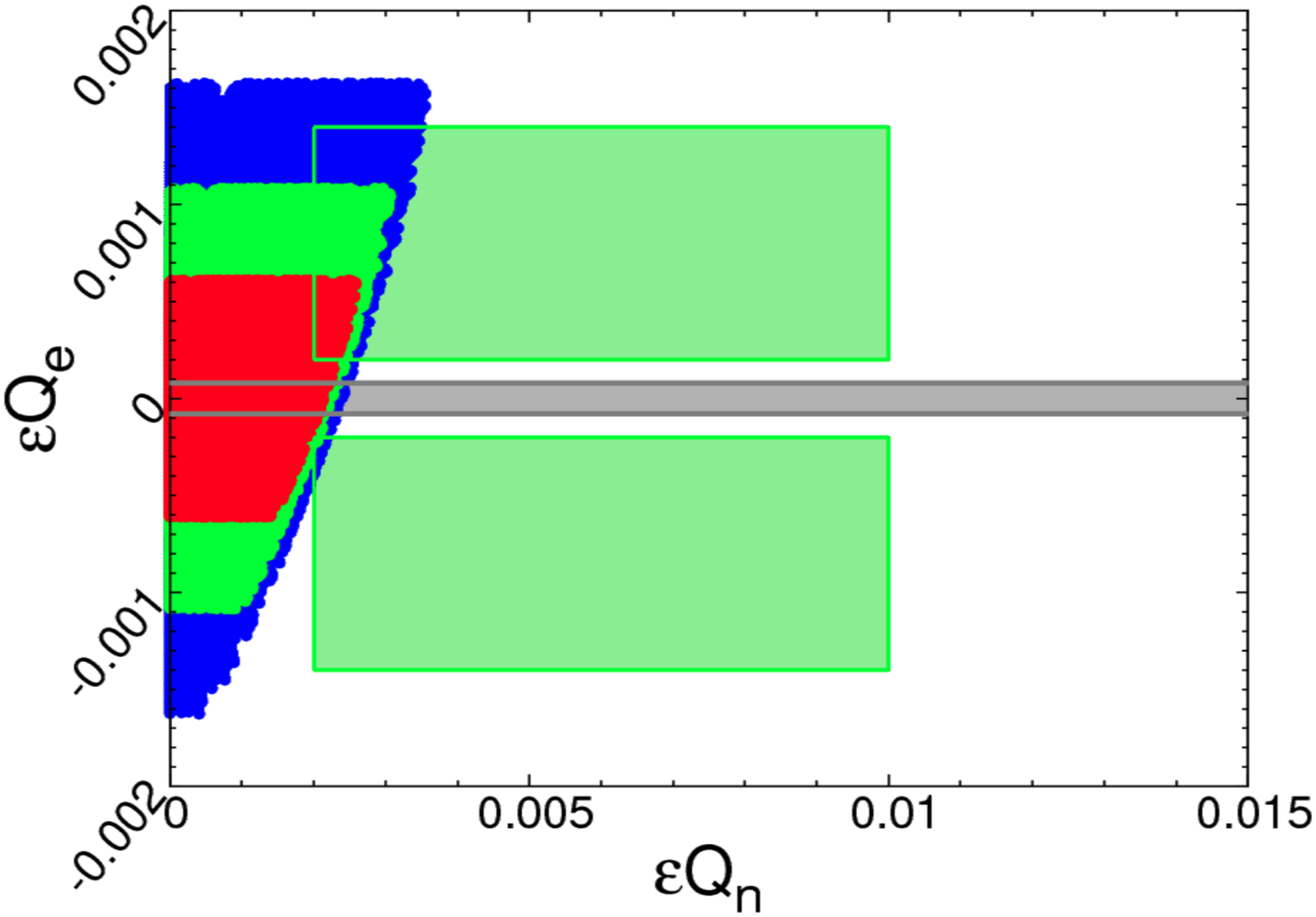}
\includegraphics[height=1.5in,angle=0]{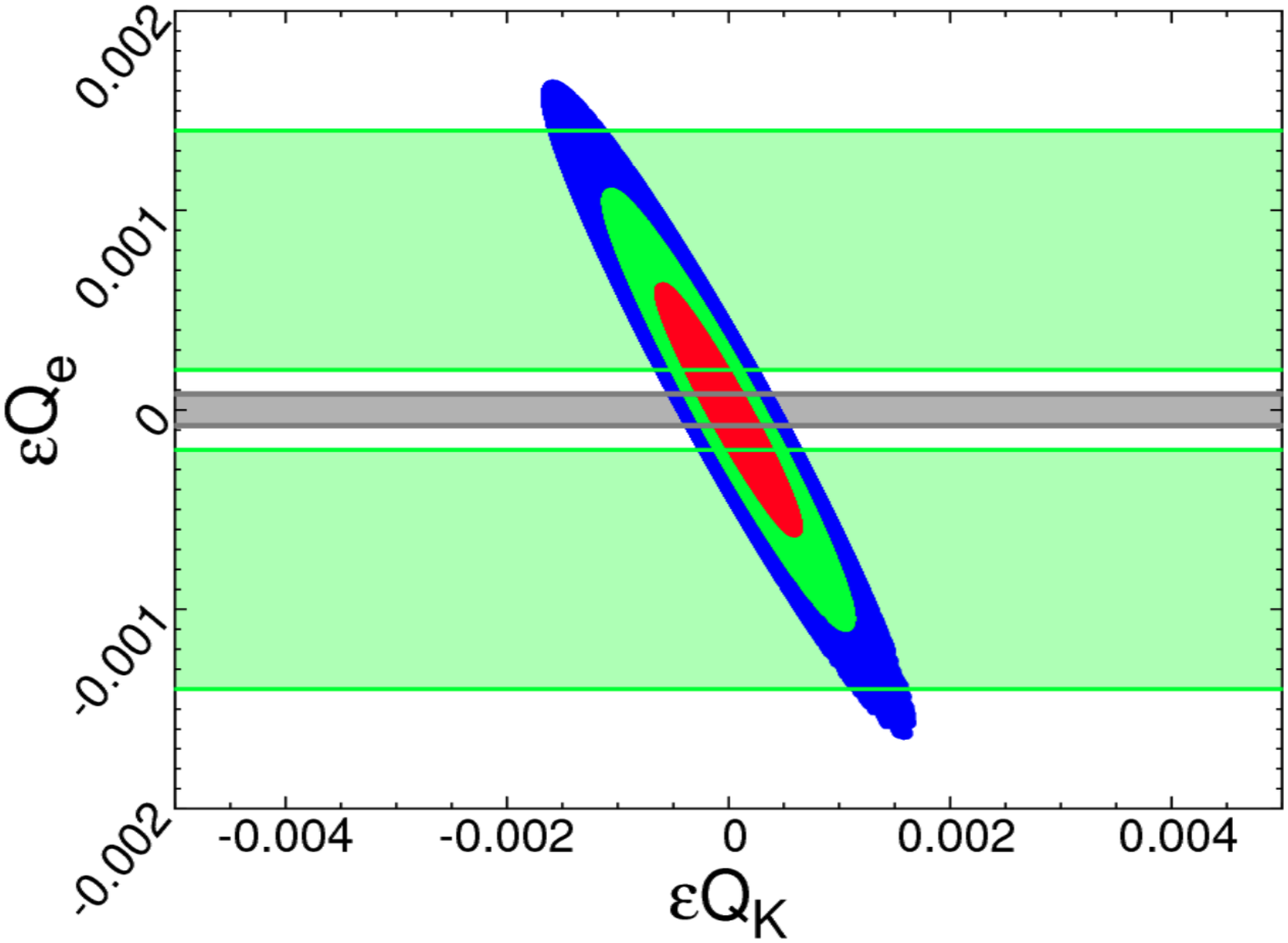}
\includegraphics[height=1.5in,angle=0]{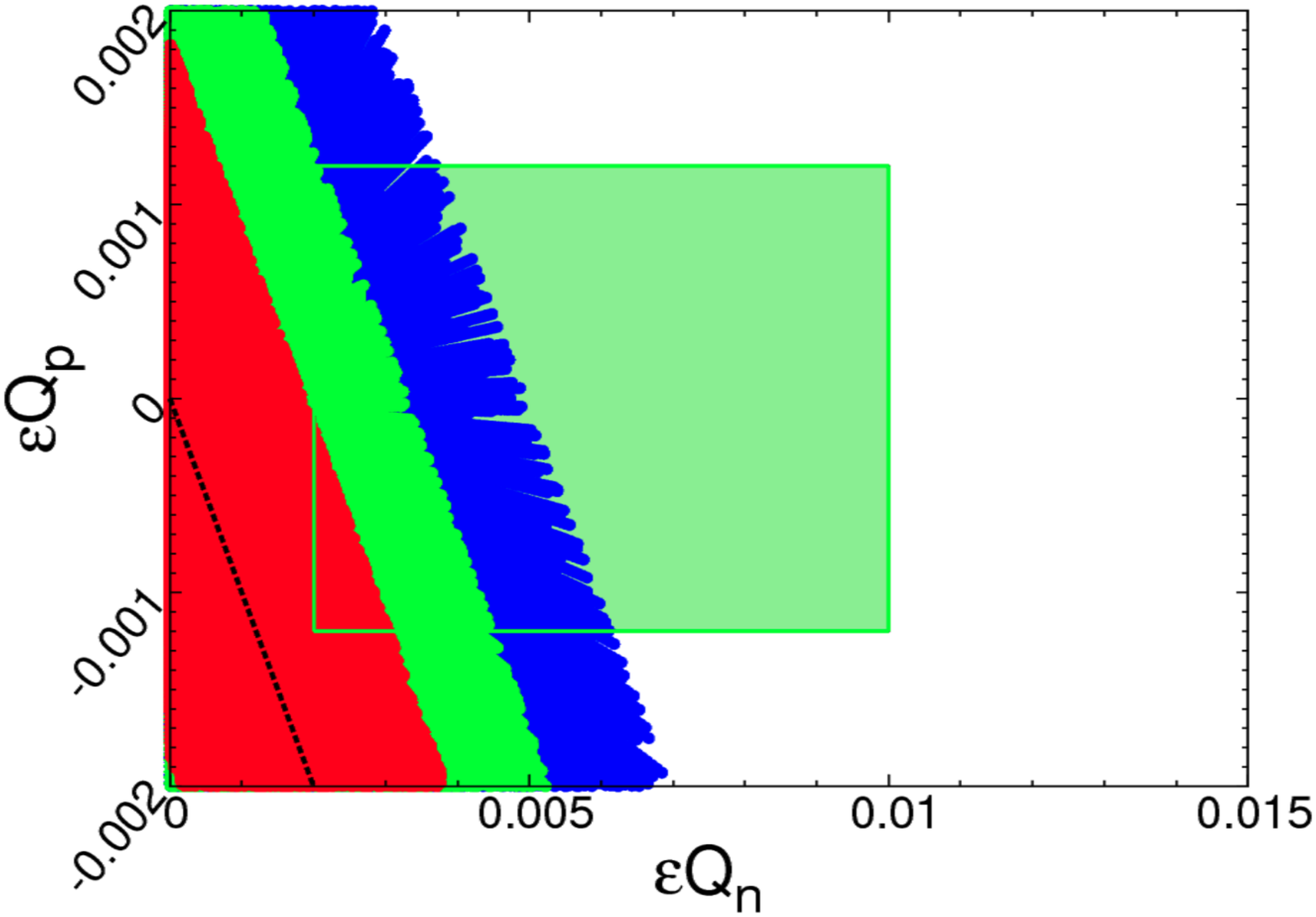}
\includegraphics[height=1.5in,angle=0]{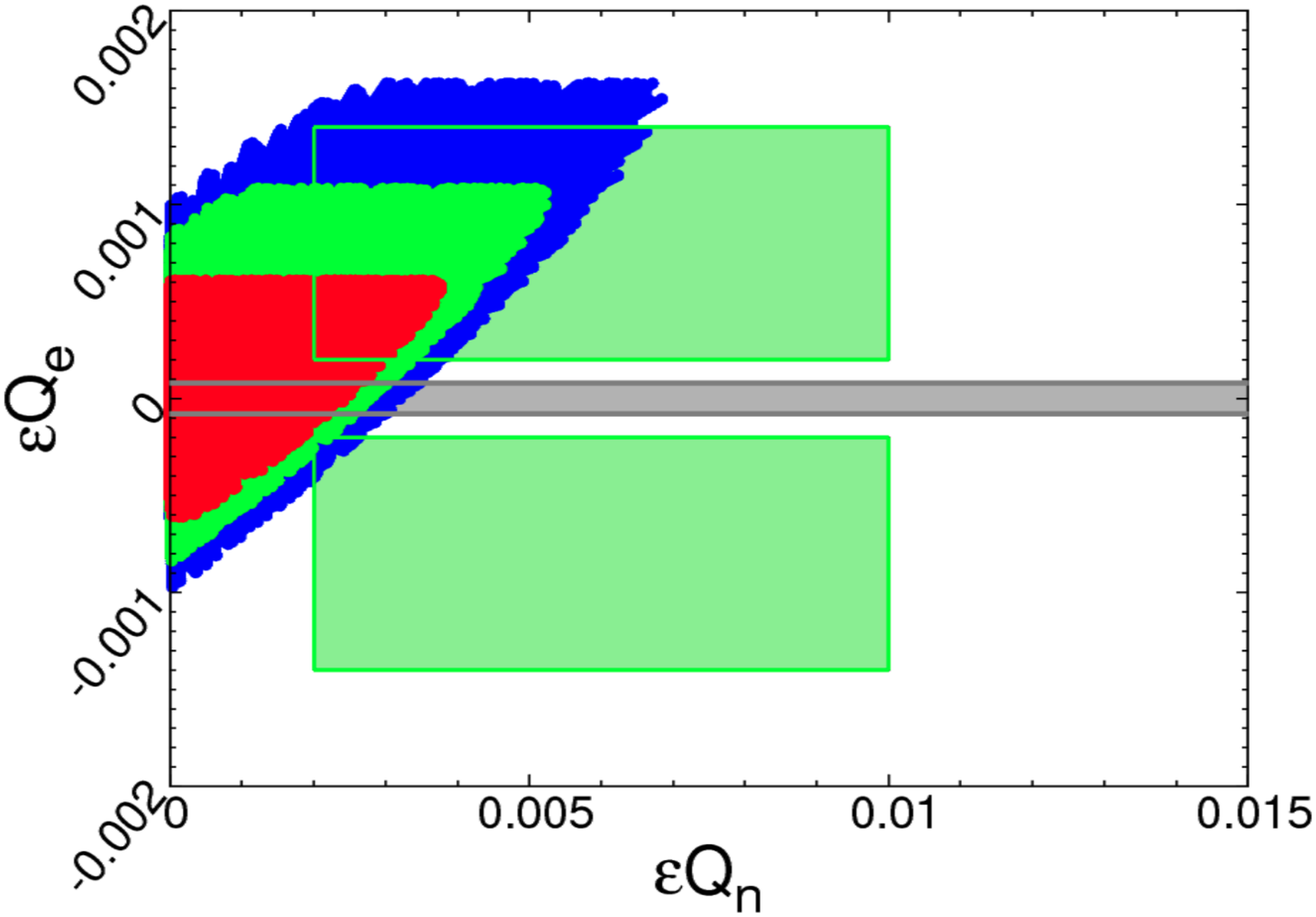}
\includegraphics[height=1.5in,angle=0]{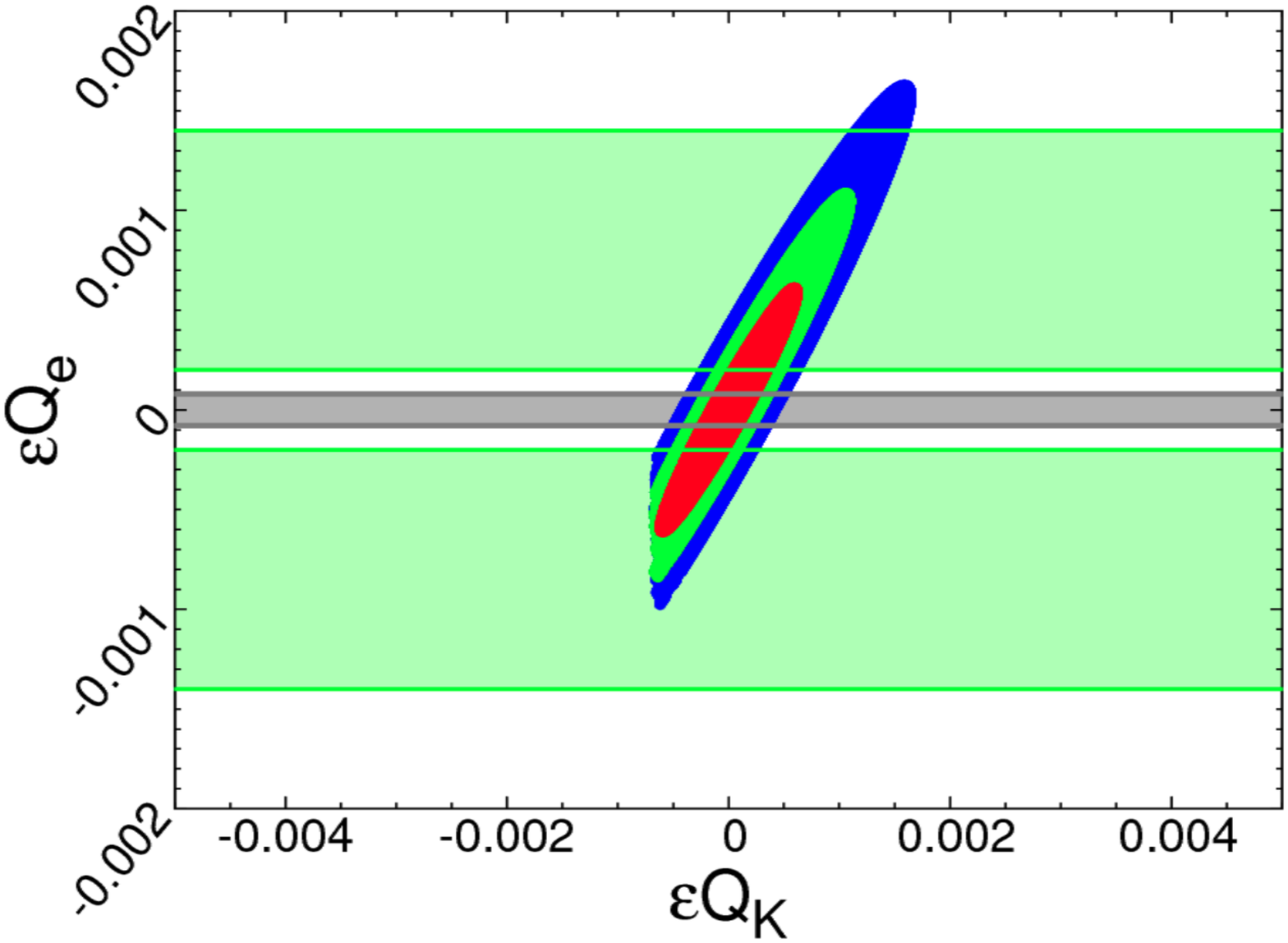}
\caption{\small \label{f3} 
The 2$\sigma$ projected limits from the $K^+ \to \mu^+ \nu e^+ e^-$ decay under the experimental Scheme 1 (red), Scheme 2 (green), and Scheme 3 (blue) and for Scenario 1 (first row), Scenario 2 (second row), and Scenario 3 (third row).  The light green area is the coupling region favored by the $^8$Be anomaly~\cite{jonathan_1,jonathan_2}.  Depending on the schemes, the region outside the red, green, or blue areas are experimentally probe-able.  The back dotted lines in the left-column plots indicate the case with $Q_{K^+}=0$.  The grey regions in the middle-and right-column plots indicate the possibility that the $X$ has a decay length larger than 1~cm. 
}
\end{figure}
\begin{figure}[h]
\includegraphics[height=1.5in,angle=0]{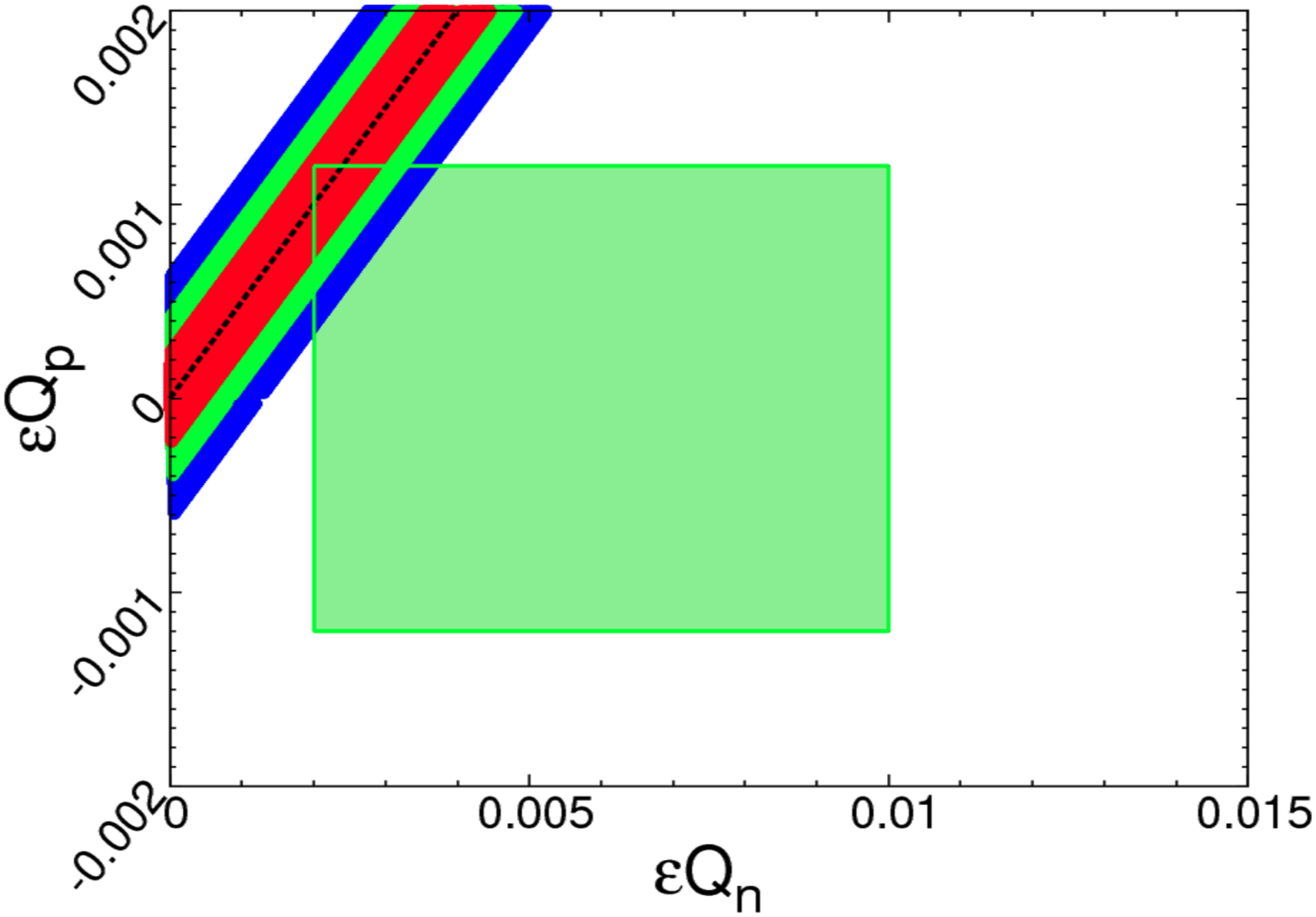}
\includegraphics[height=1.5in,angle=0]{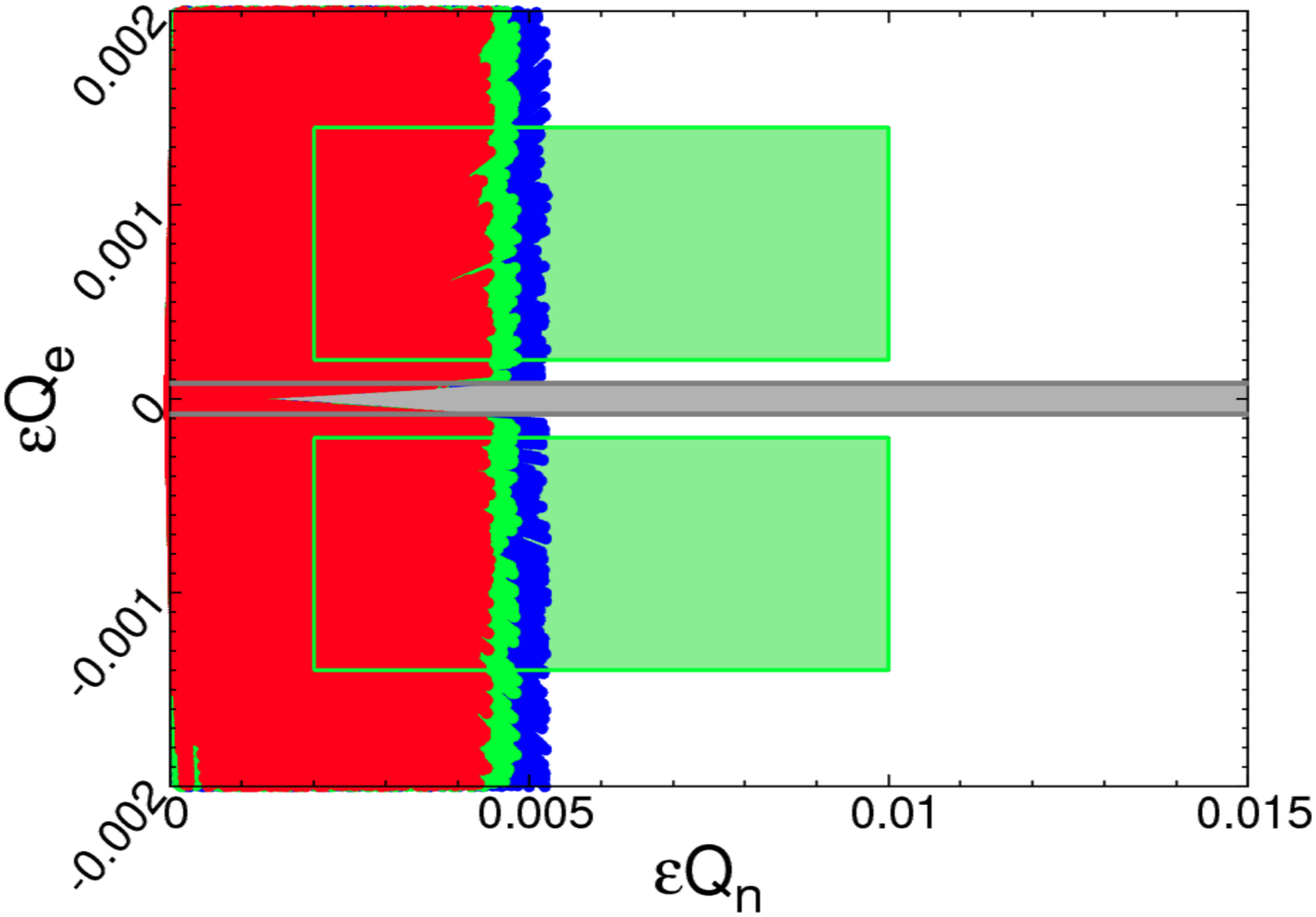}
\includegraphics[height=1.5in,angle=0]{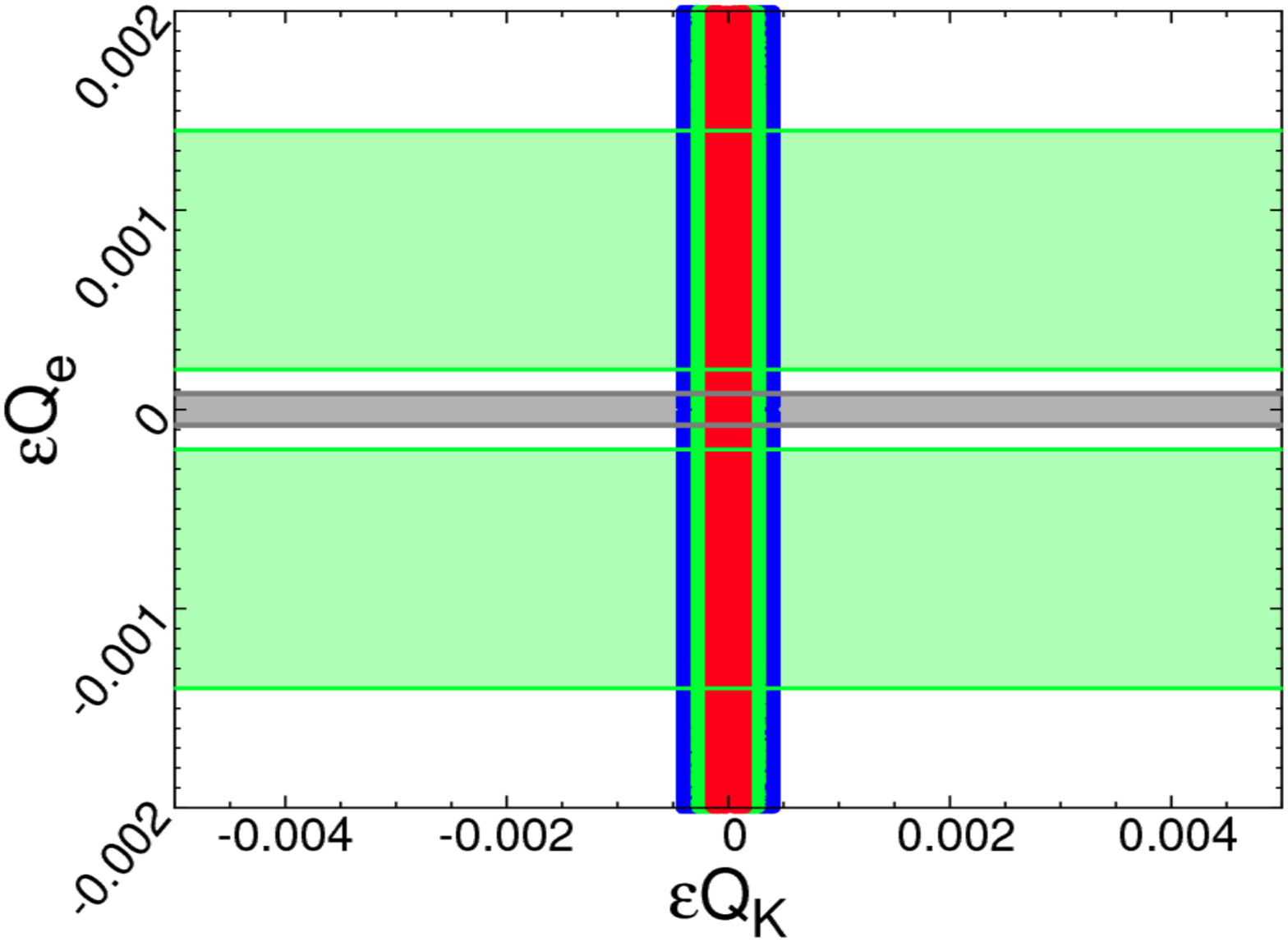}
\includegraphics[height=1.5in,angle=0]{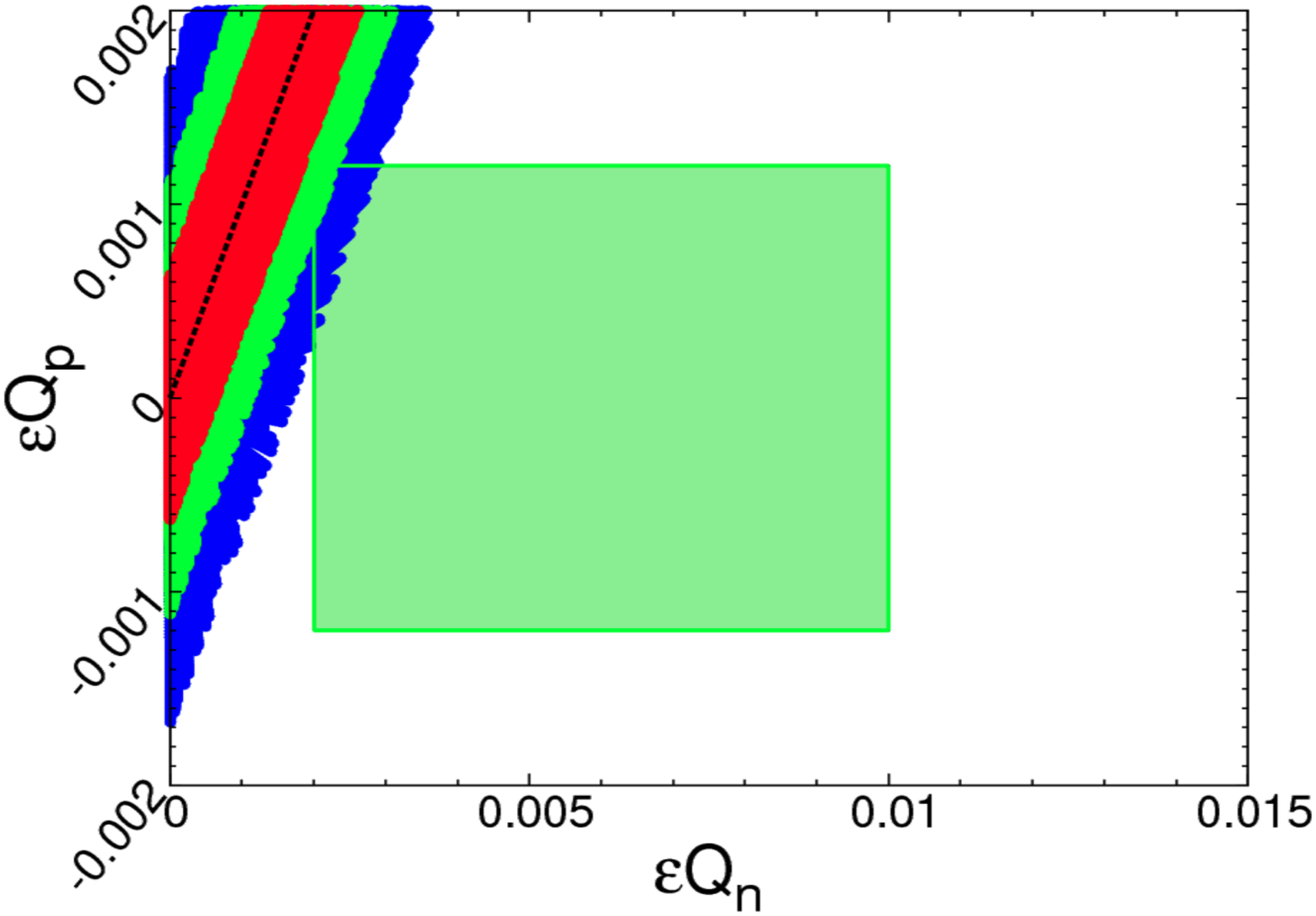}
\includegraphics[height=1.5in,angle=0]{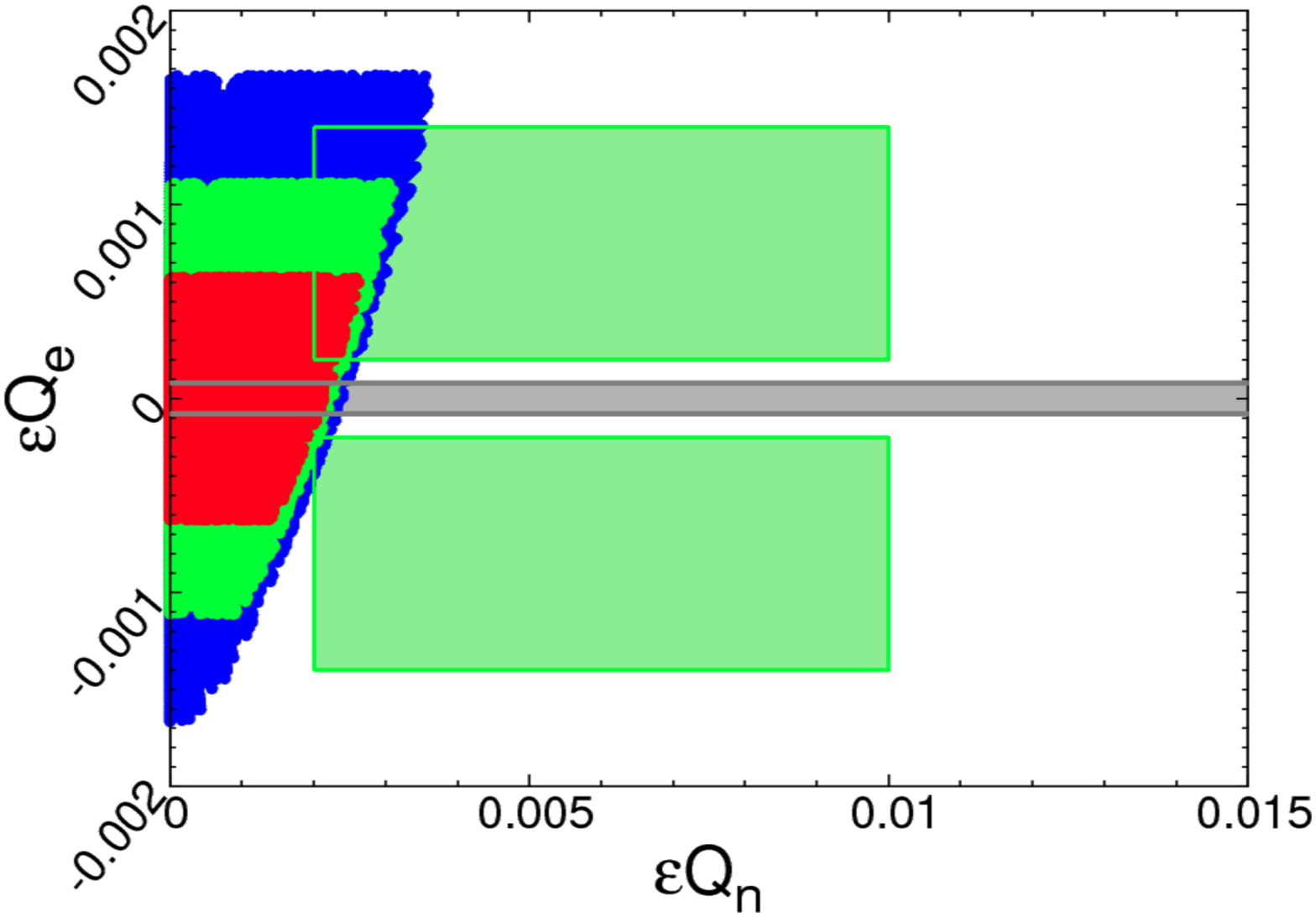}
\includegraphics[height=1.5in,angle=0]{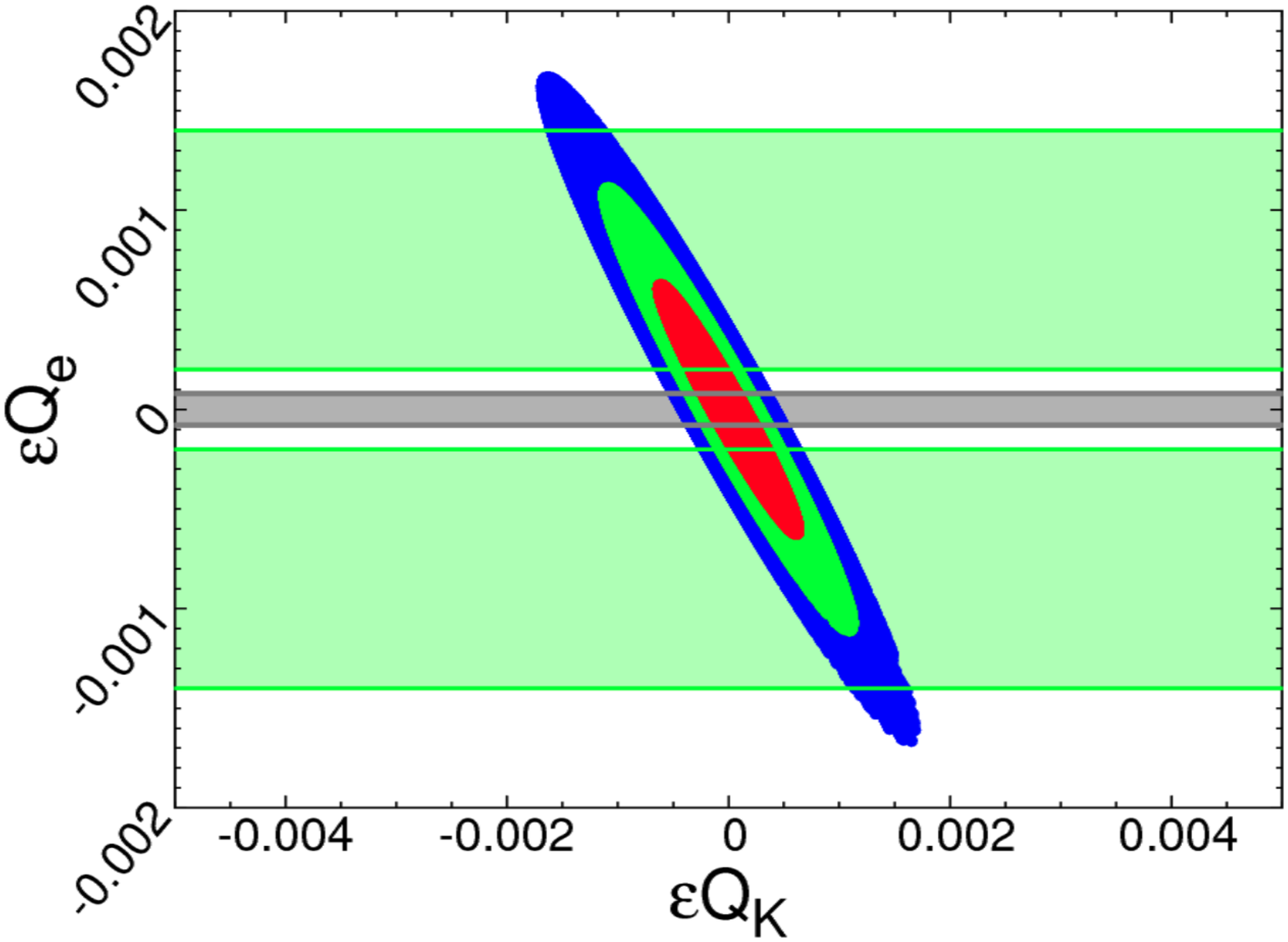}
\includegraphics[height=1.5in,angle=0]{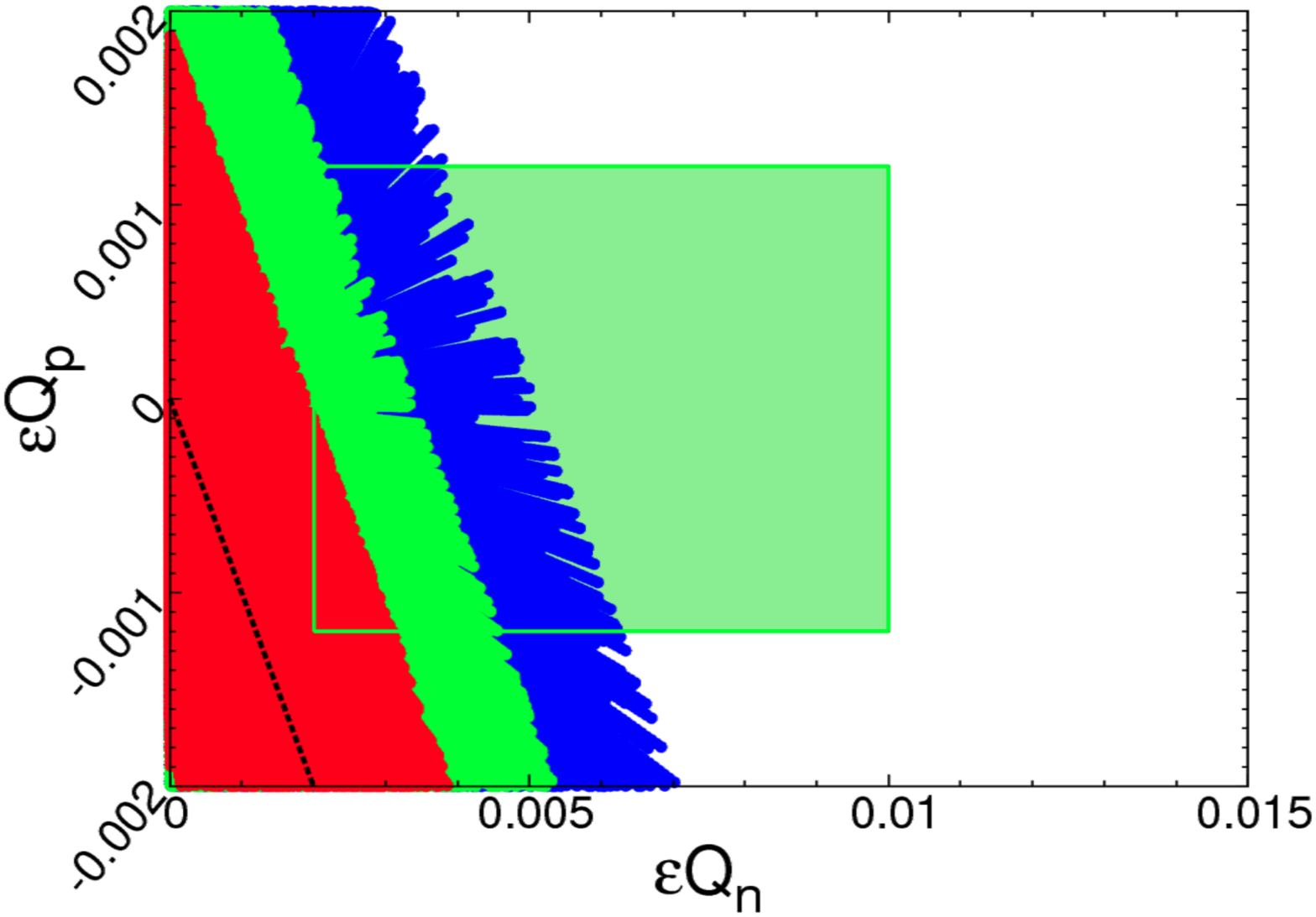}
\includegraphics[height=1.5in,angle=0]{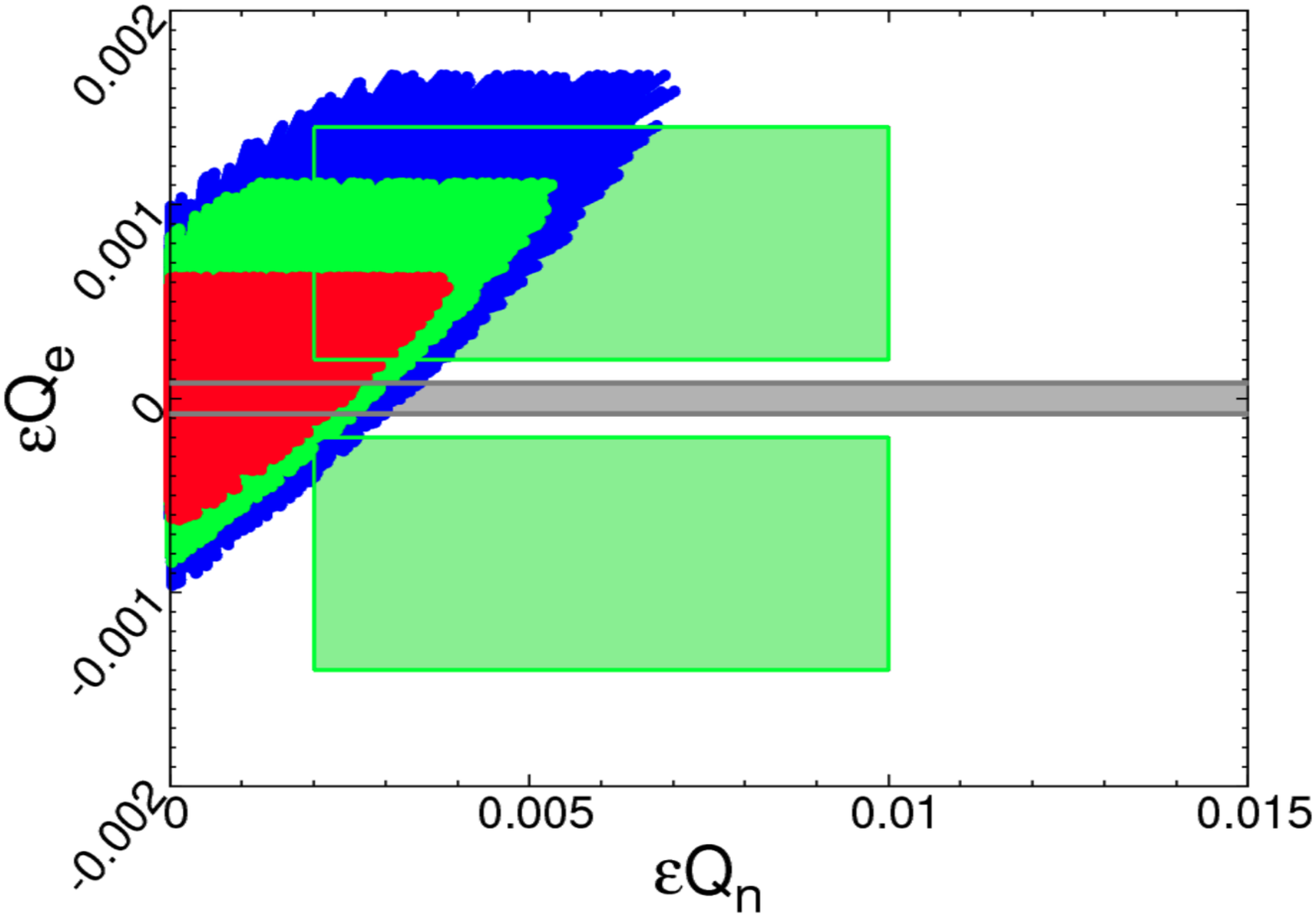}
\includegraphics[height=1.5in,angle=0]{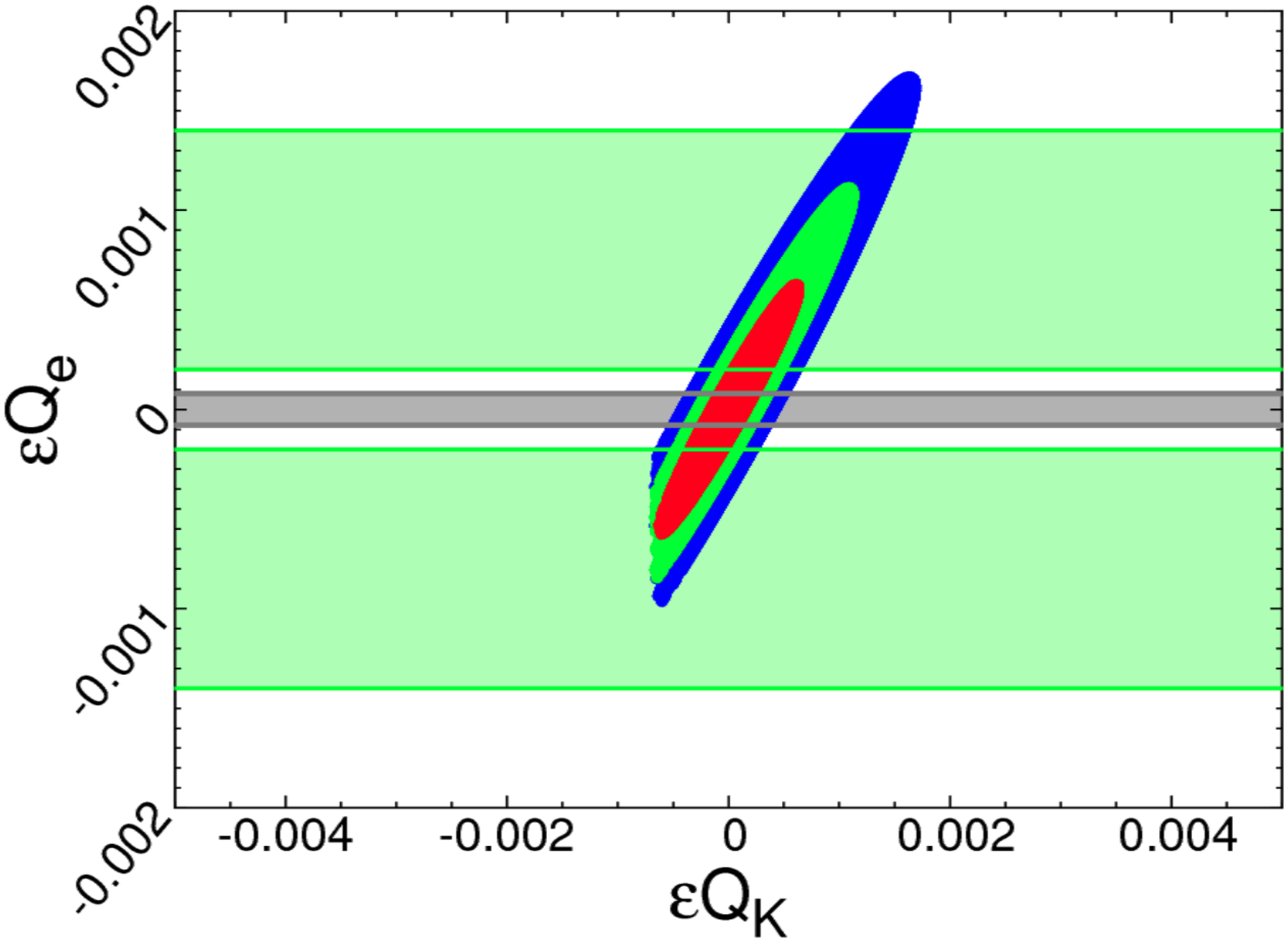}
\caption{\small \label{f4} 
Same as Fig.~\ref{f4}, but with only the IB contribution included.
}
\end{figure}
\begin{figure}[h]
\includegraphics[height=1.5in,angle=0]{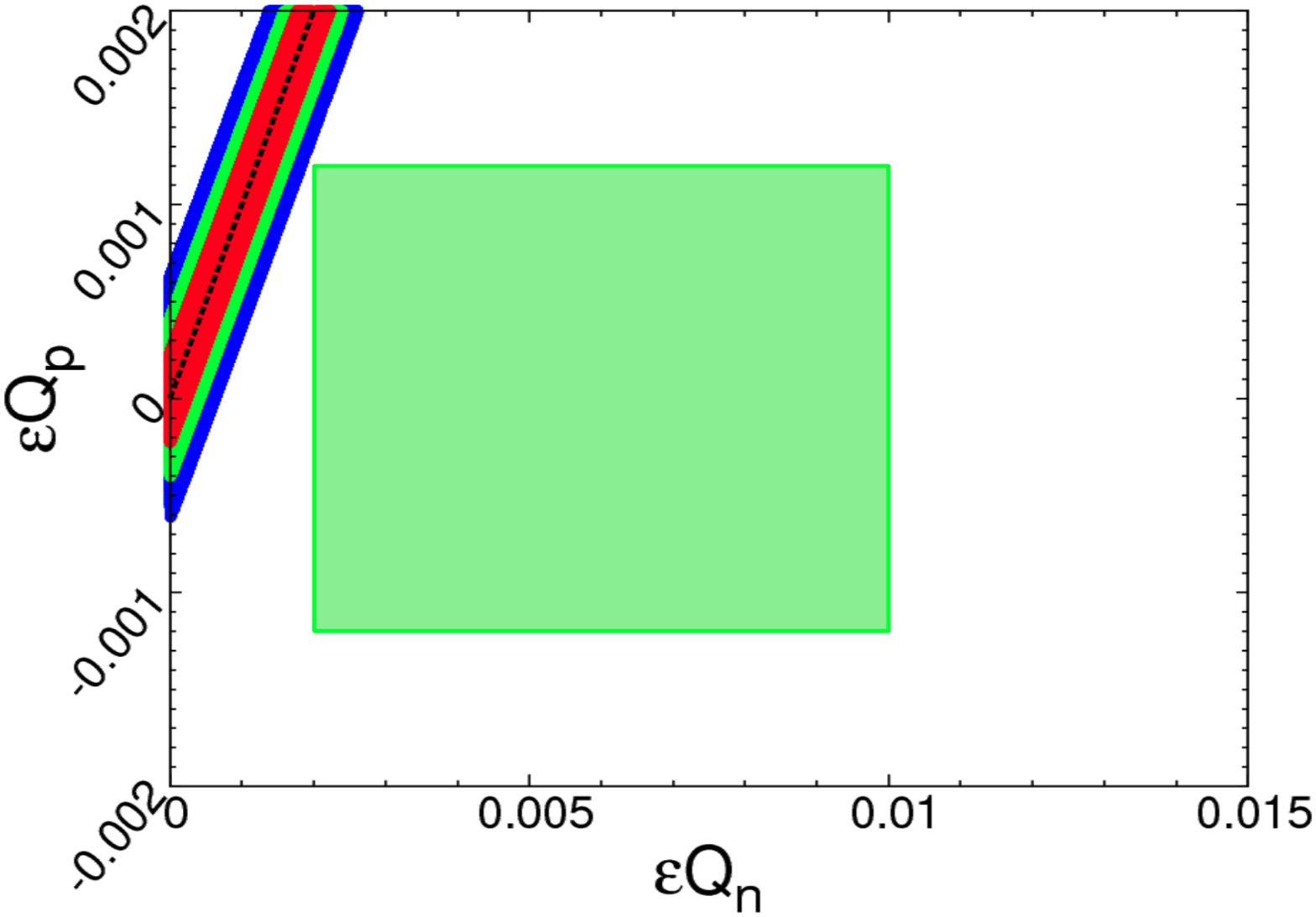}
\includegraphics[height=1.5in,angle=0]{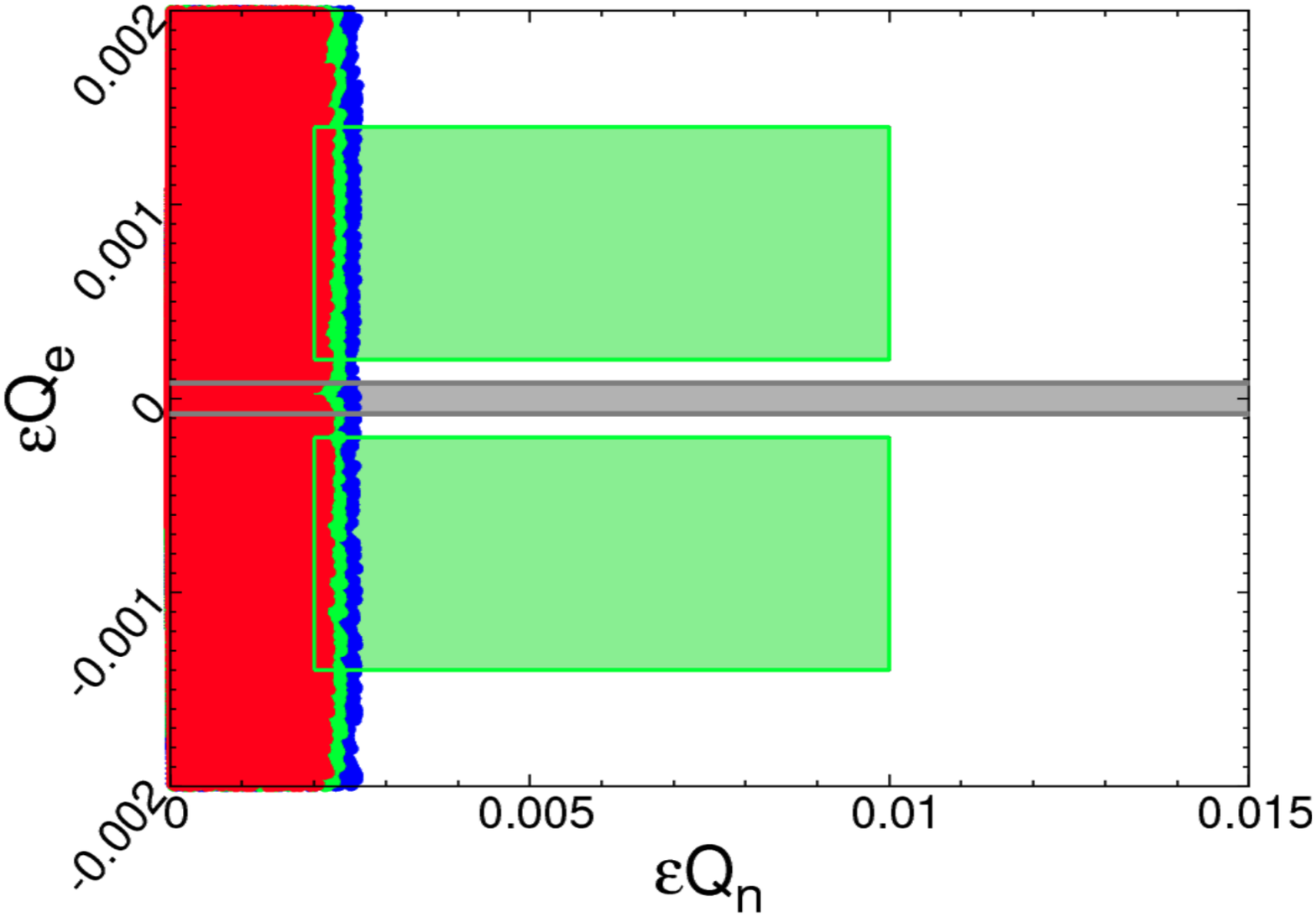}
\includegraphics[height=1.5in,angle=0]{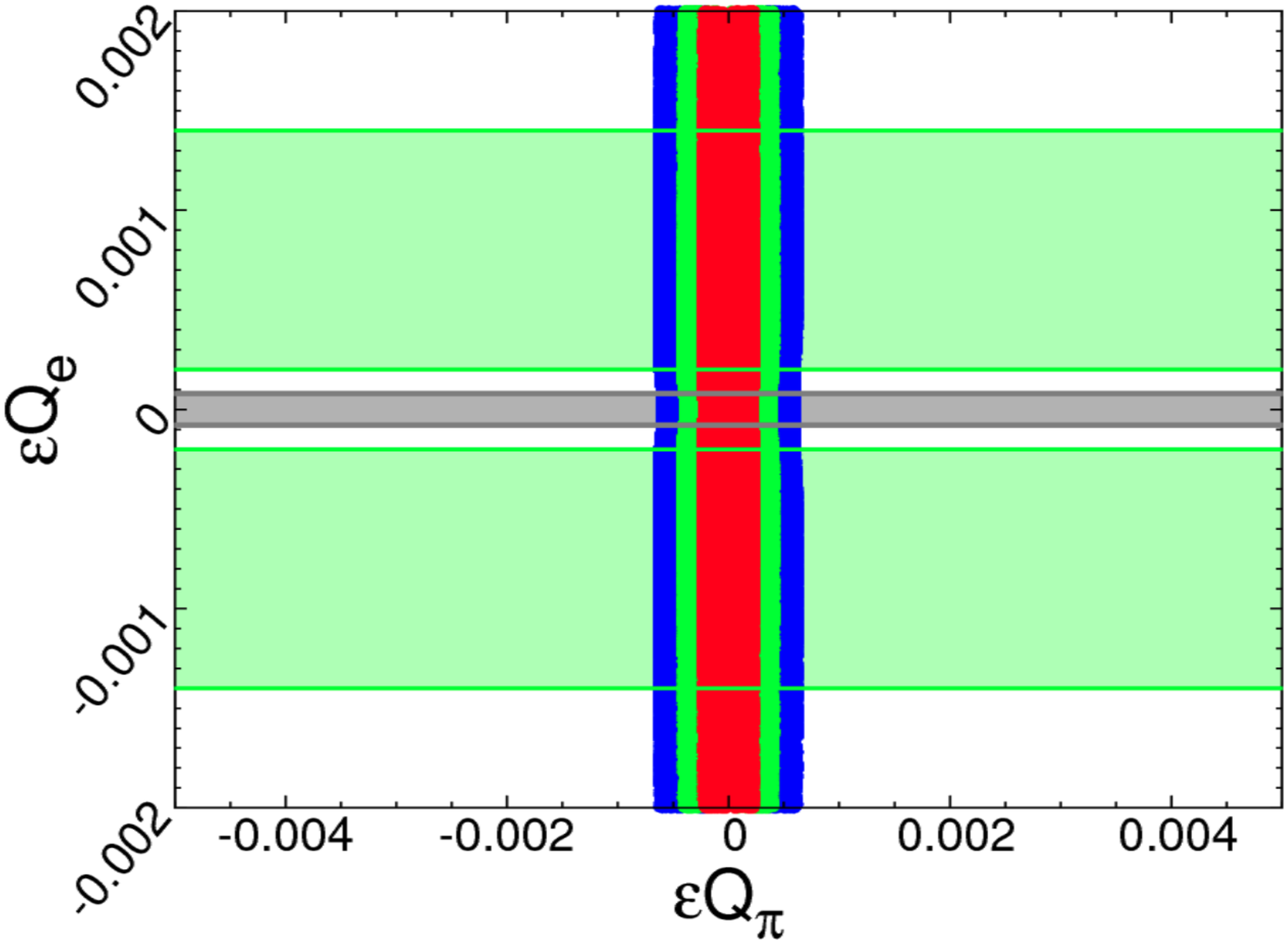}
\includegraphics[height=1.5in,angle=0]{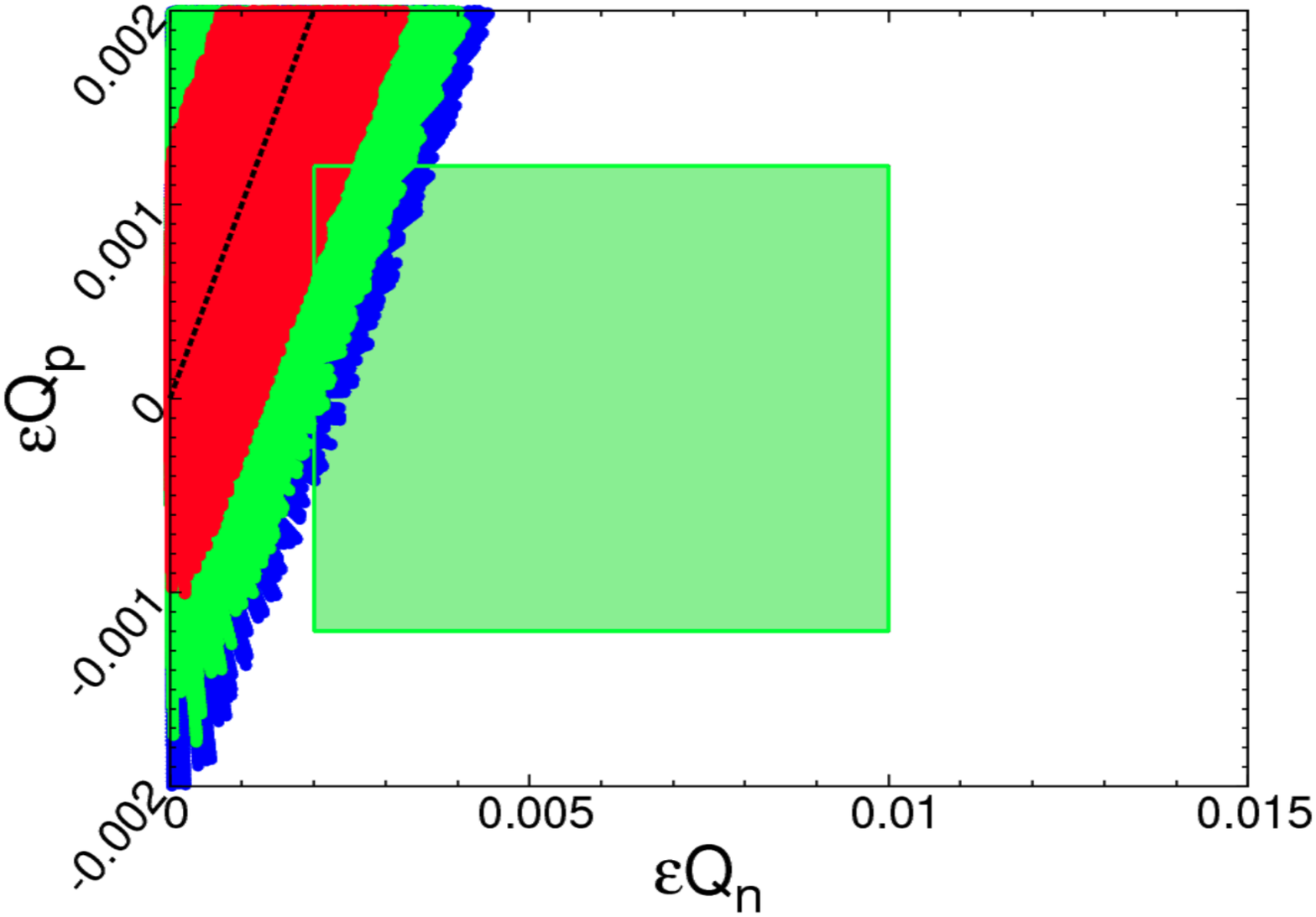}
\includegraphics[height=1.5in,angle=0]{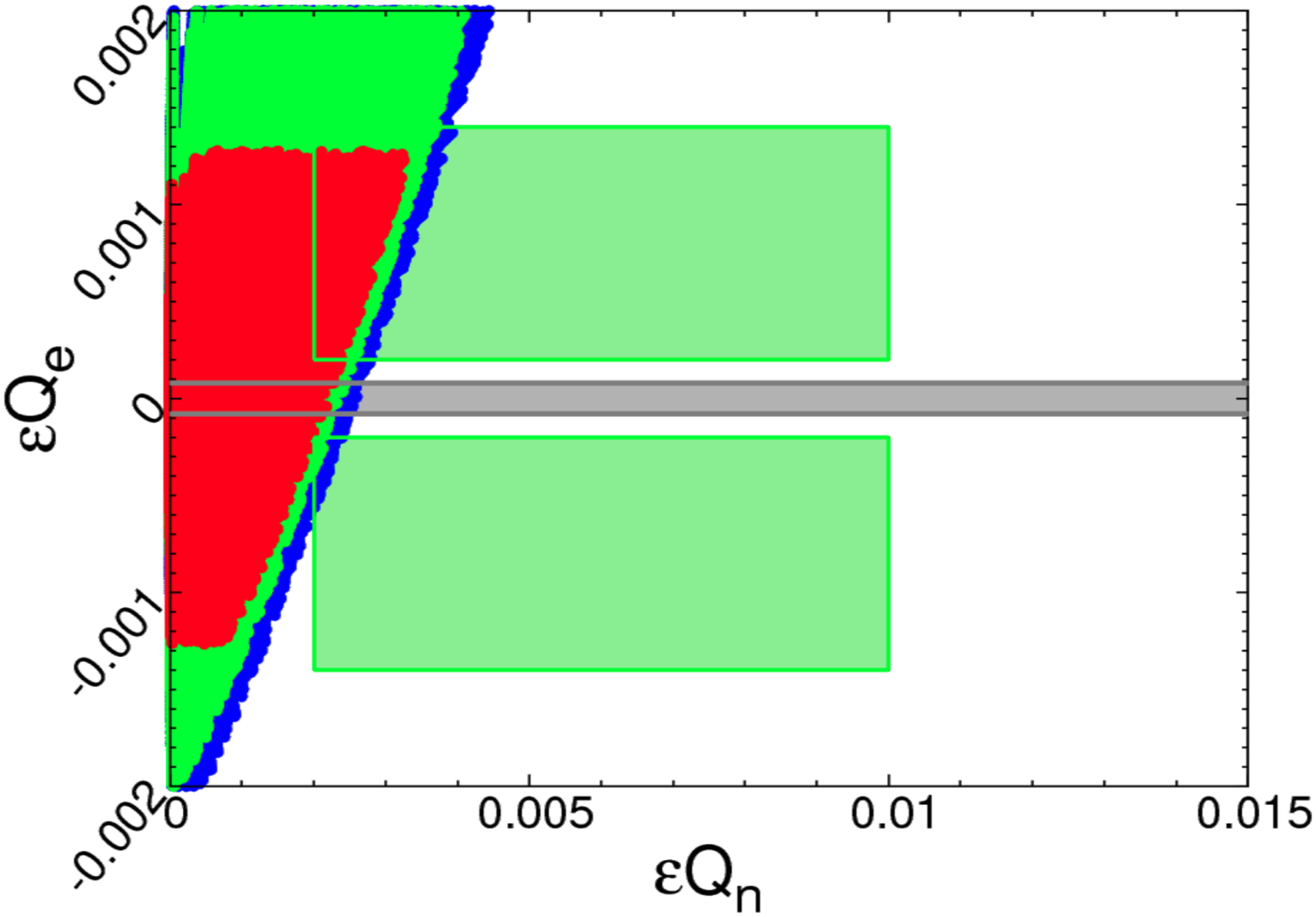}
\includegraphics[height=1.5in,angle=0]{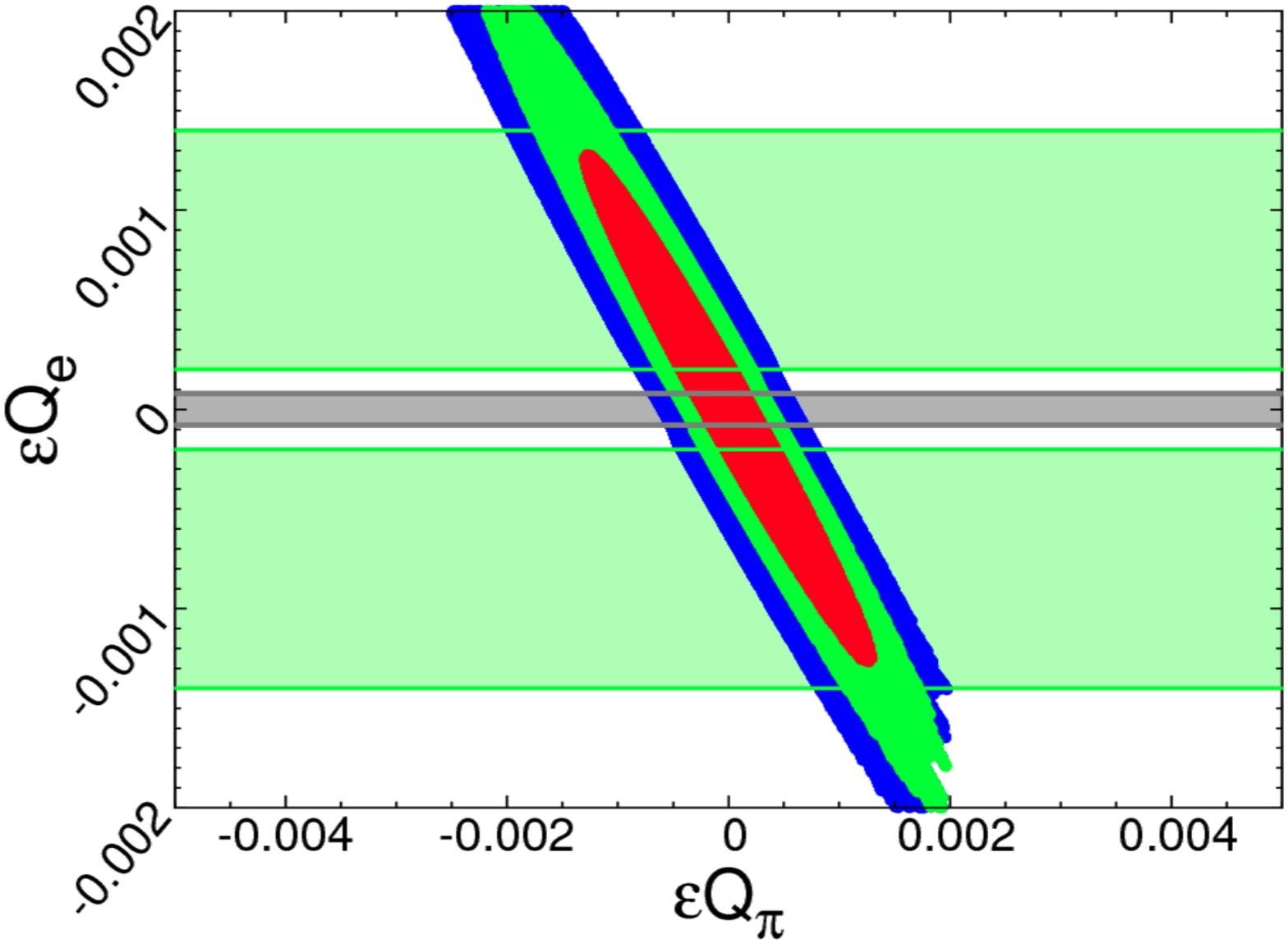}
\includegraphics[height=1.5in,angle=0]{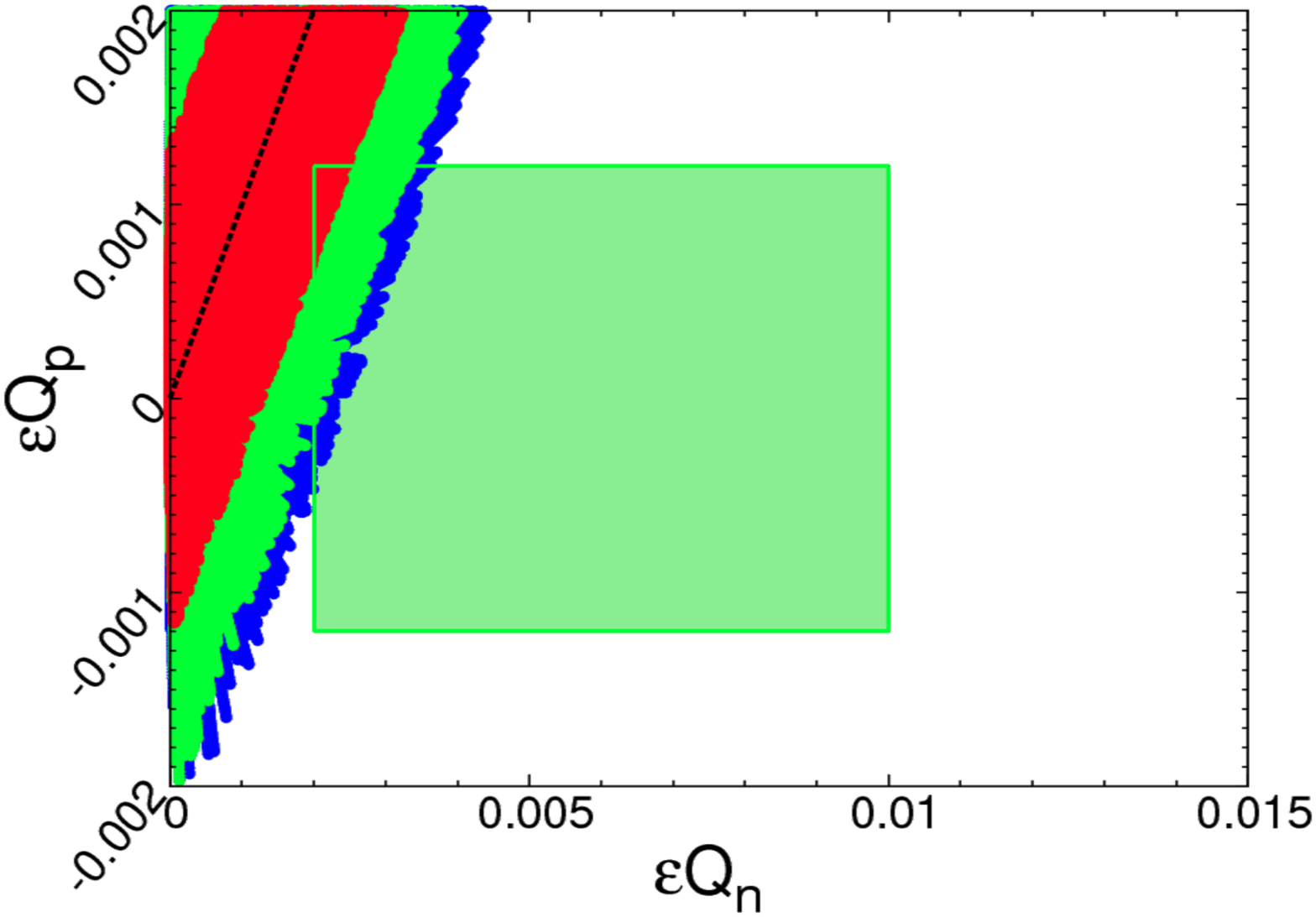}
\includegraphics[height=1.5in,angle=0]{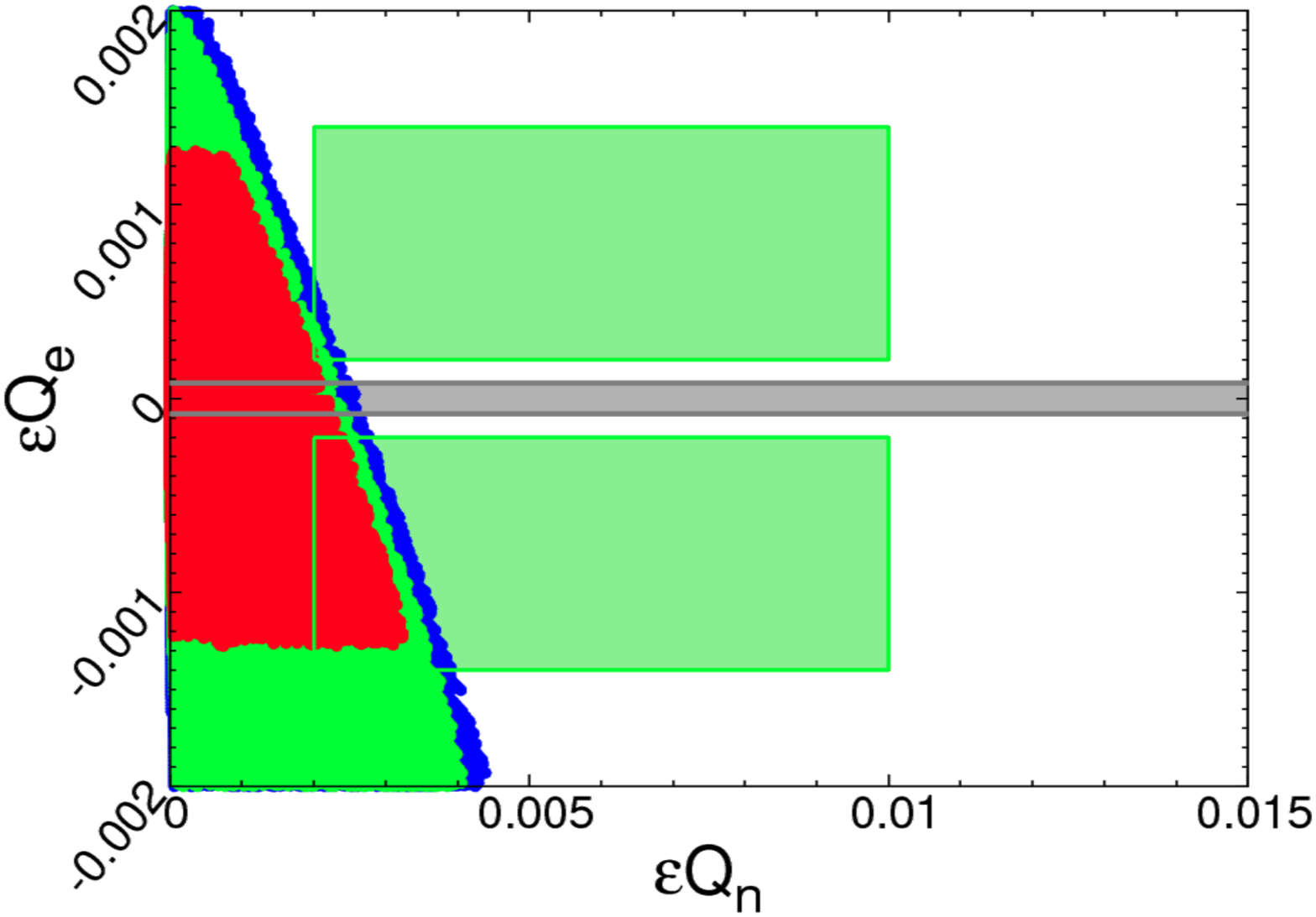}
\includegraphics[height=1.5in,angle=0]{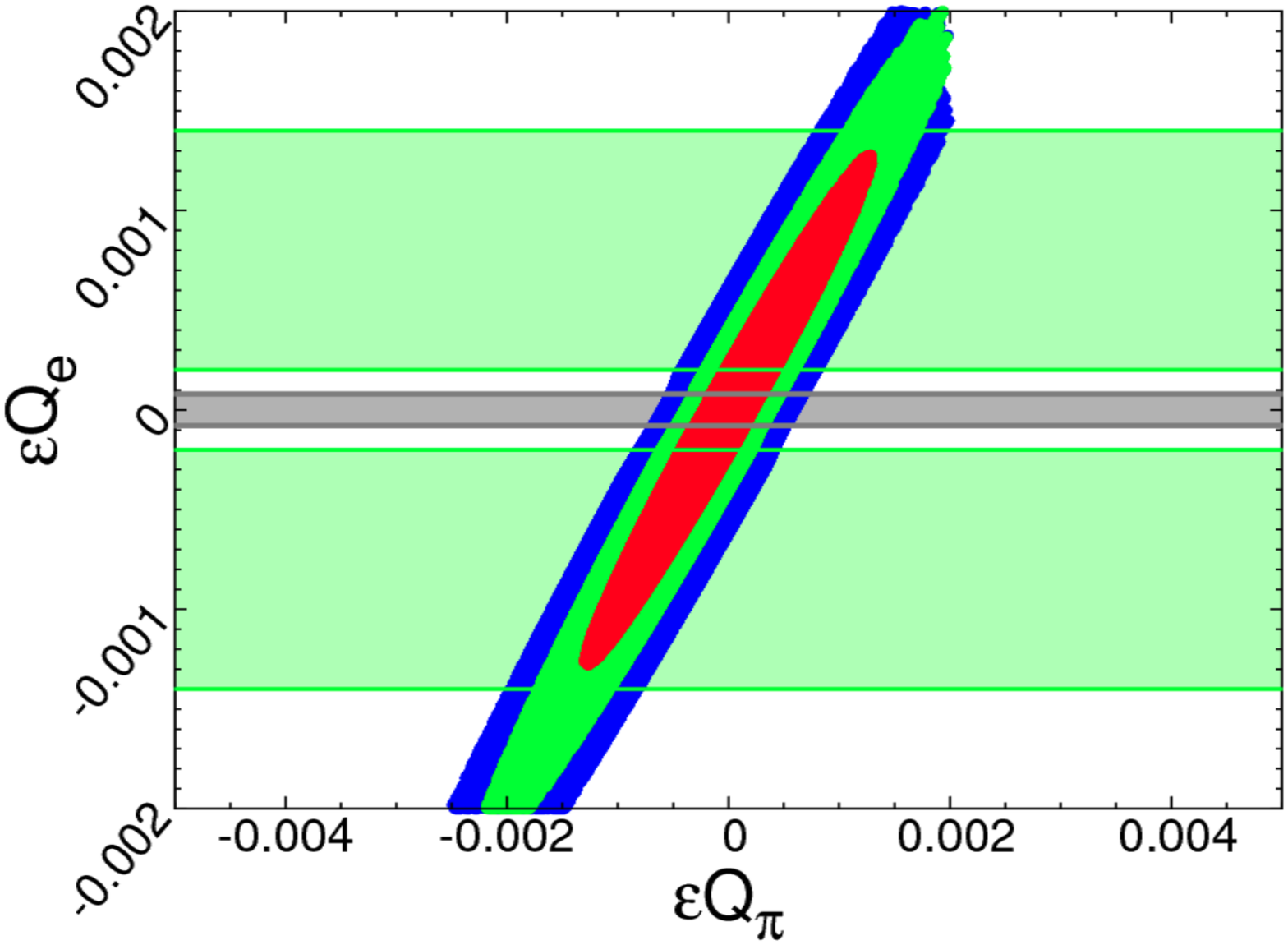}
\caption{\small \label{f5} 
Same as Fig.~\ref{f3}, but for the $\pi^+ \to \mu^+ \nu e^+ e^-$ decay.
The back dotted lines of the left-column plots indicate the case with $Q_{\pi^+}=0$.
}
\end{figure}

Since the best-fit mass of dark photon for the $^8$Be anomaly is 16.7~MeV~\cite{atomki}, we will fix $m_X=16.7$~MeV in the following discussions.  Refs.~\cite{jonathan_1,jonathan_2} worked out the ranges of dark photon couplings with neutron, proton, and electron that could explain the $^8$Be anomaly and evade other experimental constraints.  We will show how the leptonic kaon and pion decays can probe this region.

First, we take the experimental Scheme 1 and consider the kaon decay in Scenario 1 as an example.  The SM background branching ratio is $\Delta {\rm BR}_{\gamma^*}=2.54\times 10^{-7}$.  Then the signal branching ratio corresponding to $2\sigma$ upper limit of $\varepsilon_{\rm limit}$ is
\begin{equation}
{\rm BR}_{\rm limit} \equiv
\varepsilon^2_{\rm limit}\times 
\frac{{\rm BR}_{X}(\tilde{\varepsilon})}{\tilde{\varepsilon}^2}
=1.009\times 10^{-9}\,.
\end{equation}
Using ${\rm BR}_{\rm limit}$ as a basic unit, we draw the contours of ${\rm BR}_{X}$ on the $(\varepsilon Q_n, \varepsilon Q_p)$ parameter plane in Fig.~\ref{f2}.  The black dotted line in the plot indicates $\varepsilon Q_{K^+}=0$ and thus ${\rm BR}_{X}=0$.  The region with ${\rm BR}_{X} < {\rm BR}_{limit}$ is between the two red lines.  This is the region where the dark photon in Scenario 1 cannot be checked using the leptonic kaon decay under the experimental Scheme 1.  However, all the region outside the red contour, including most of the light green region, can be probed.

We perform a scan over the dark photon couplings, $(\varepsilon Q_e,\varepsilon Q_p, \varepsilon Q_n)$. In Fig.~\ref{f3}, we show the $2\sigma$ projected limits for Scenarios 1, 2, and 3 in experimental Schemes 1, 2, and 3.  Most part of the light green area, outside the red, green, or blue regions, can be probed by the $K^+ \to \mu^+ \nu e^+ e^-$ decay.  In the left column, the black dotted lines indicate the case with $Q_{K^+}=0$.  In this case, the leptonic kaon decay loses sensitivity to probing the dark photon.  In the right column, the leptonic kaon decay also has less sensitivity to the dark photon along the semi-major axis of the ellipses.  This is because when $Q_{K^+}=-Q_{\mu}$, a cancellation occurs in the IB amplitude and, therefore, ${\rm BR}_X$ is suppressed.

Fig.~\ref{f4} is the same as Fig.~\ref{f3}, except that we turn off the SD contribution in the amplitude of leptonic kaon decay.  There is almost no noticeable numerical difference between Fig.~\ref{f3} and Fig.~\ref{f4}.  This demonstrates that uncertainties from the kaon form factors have little effects on the analysis, especially for $m_X \simeq 16.7$ MeV.

In Fig.~\ref{f5}, we show the $2\sigma$ projected limits from the $\pi^+ \to \mu^+ \nu_{\mu} e^+ e^-$ decay as another possible probe.  Comparing to Fig.~\ref{f3}, the projected limits from the leptonic pion decay are not as good.  But a good portion of the preferred parameter region for the $^8$Be anomaly (light green area) can still be probed.

\begin{figure}[t!]
\includegraphics[height=3.5in,angle=0]{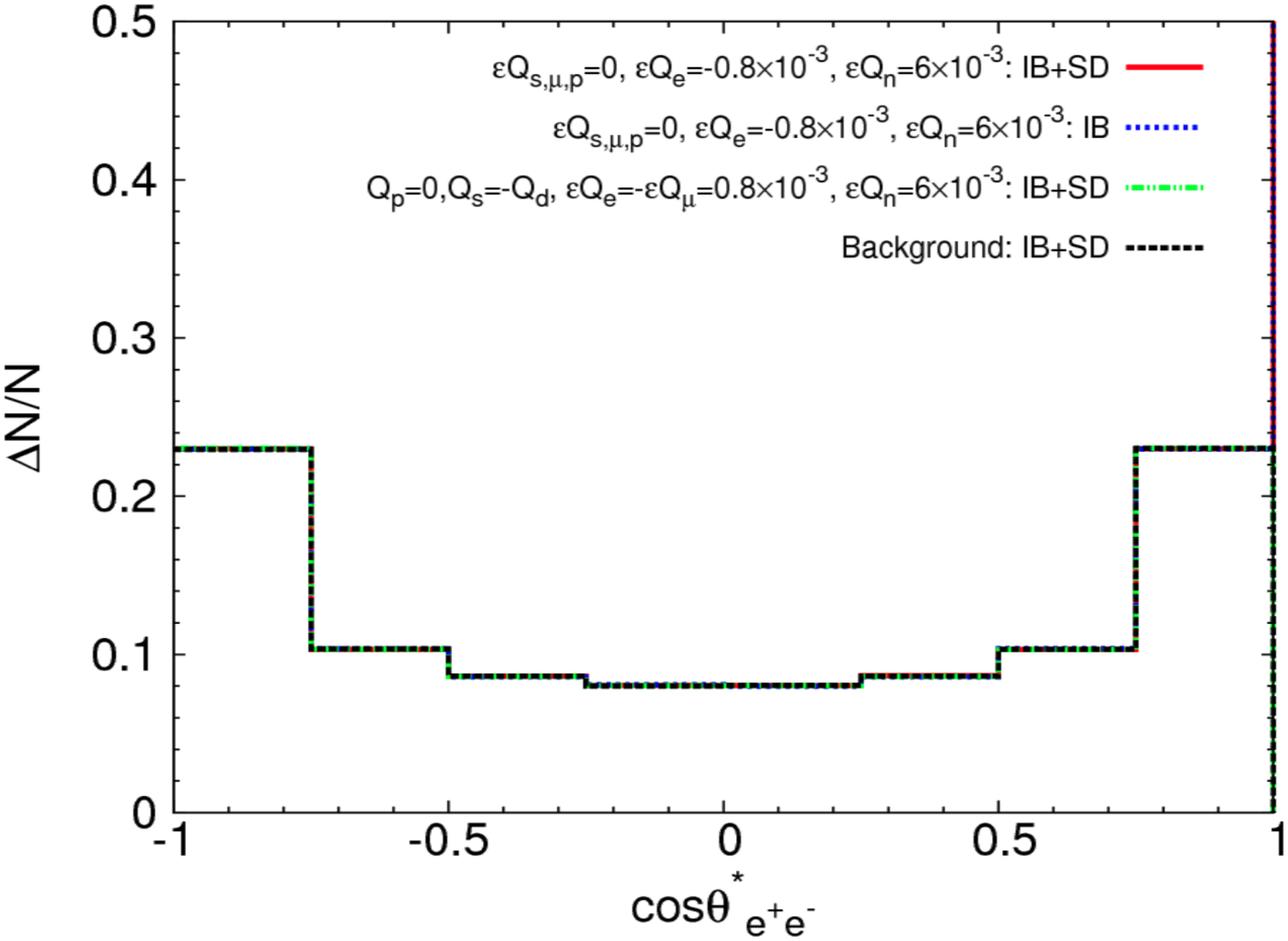}
\caption{\small \label{f6} 
Angular distribution of the $K^+ \to \mu^+ \nu_{\mu} e^+ e^-$ decays, where $\theta^*_{e^+e^-}$ denotes the angle of $e^+$ with respect to the 3-momentum of $X$, but measured in the $e^+e^-$ center-of-mass frame.
The $\Delta {\rm N}/{\rm N}$ is the ratio between the number of events within each bin to the total number of events.  Virtually no difference can be seen among the four scenarios.
}
\end{figure}
\begin{figure}[t!]
\includegraphics[height=2.2in,angle=0]{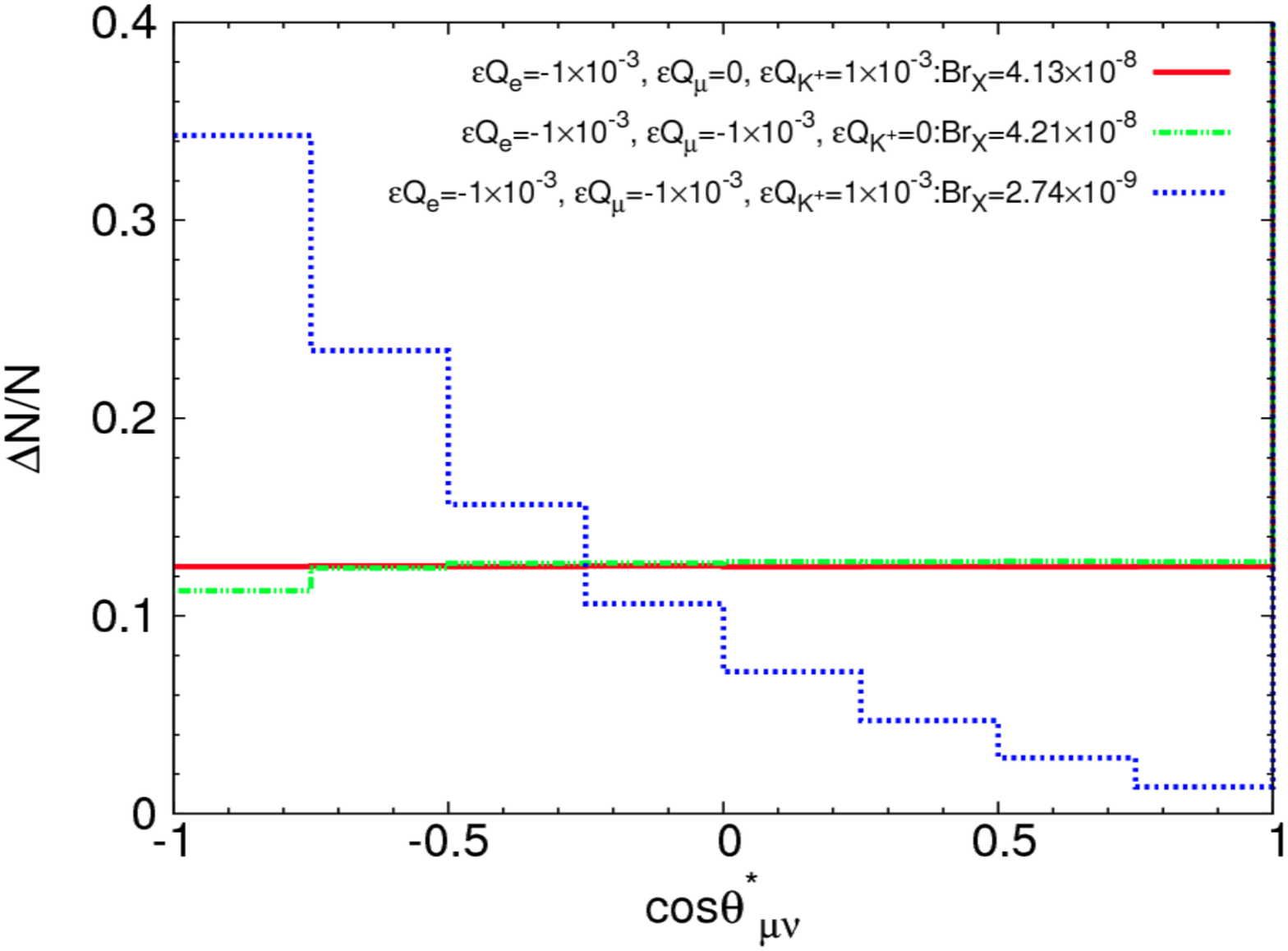}
\includegraphics[height=2.2in,angle=0]{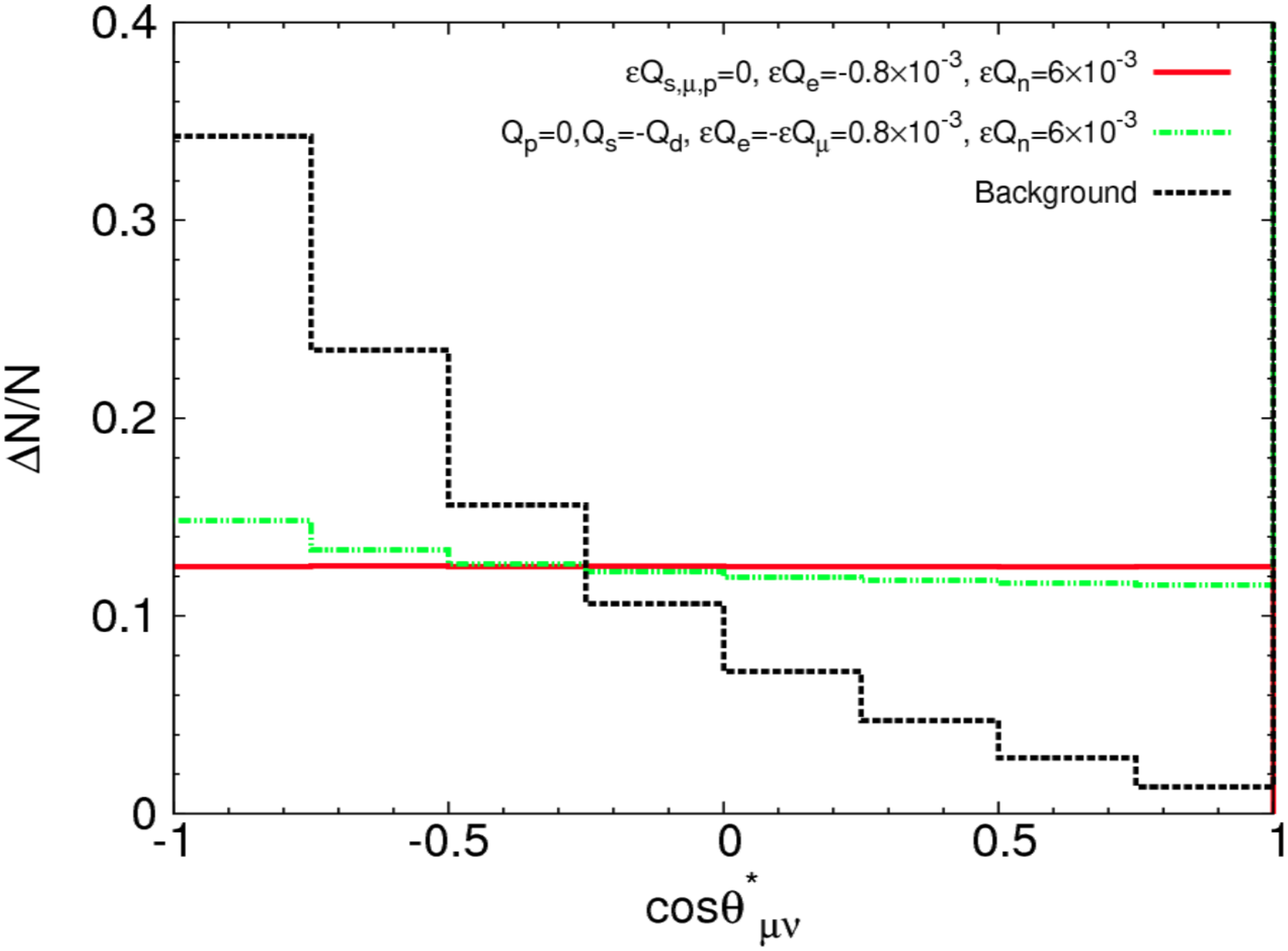}
\caption{\small \label{f7} 
The $\cos\theta^*_{\mu\nu}$ distribution of the $K^+ \to \mu^+ \nu_{\mu} e^+ e^-$ decays, where $\theta^*_{\mu\nu}$ denotes the angle of $\mu^+$ with respect to the $\mu^+\nu$ 3-momentum sum, but measured in the $\mu^+\nu$ center-of-mass frame.  Both IB and SD contributions are included in the calculations.
}
\end{figure}

We now turn to the discussions of angular distributions of different subsystems in the $\mu^+ \nu_{\mu} e^+ e^-$ final state.
First, we show in Fig.~\ref{f6} the distribution of $\cos\theta^*_{e^+e^-}$ from the leptonic kaon decay, where $\theta^*_{e^+e^-}$ denotes the angle of $e^+$ with respect to the boost direction of $X$, but measured in the $e^+e^-$ center-of-mass frame.  The parabolic shape of the angular distribution signifies the feature of vectorial couplings of the $X$ boson and the photon, and it is almost independent of $U(1)'$ charges and the SD part.

In Fig.~\ref{f7}, we show the distribution of $\cos\theta^*_{\mu\nu}$, where $\theta^*_{\mu\nu}$ is the angle of $\mu^+$ with respect to the boost direction of the $\mu^+\nu$ system, but measured in the $\mu^+\nu$ center-of-mass frame.  Apparently, the SM background and signal from dark photon have different behaviours.
From the SM background, the $\mu^+$ tends to fly toward the direction of $X$ in the $K^+$ rest frame.
For the signal events, however, it has an isotropic distribution.
In the left plot, when $\varepsilon Q_{K^+}\neq 0$ and $\varepsilon Q_{\mu}=0$, we are left with only the first term in the curly brackets of $\bar{L}^{\rho}$ in Eq.~(\ref{eq2}) that has no angular dependence.
When $\varepsilon Q_{K^+}=0$ and $\varepsilon Q_{\mu}\neq 0$, on the other hand, the second term in the curly brackets survives and has some minor angular dependence. 
When $\varepsilon Q_{K^+}=-\varepsilon Q_{\mu}$, the angle-independent parts cancel, rendering a more dramatic angular dependence, as shown by the blue-dotted histogram.  The same cancellation also happens for the SM background, as the electric charges of $K^+$ and $\mu$ have opposite signs.  The result is shown by the black-dotted histogram in the right plot.

\section{Summary~\label{sec:summary}}

In this work, we propose to use the rare leptonic kaon and pion decays, $K^+/\pi^+ \to \mu^+\, \nu_{\mu}\, e^+ e^-$ to probe a light dark photon of ${\cal O}(10)$~MeV mass.  This is particularly suitable for probing the putative $X$ gauge boson hinted at by the recent $^8$Be anomaly.  We consider three scenarios for the standard model particle couplings with the dark photon.  The first scenario assumes that the dark photon only couples to fermions in the first family, as directly relevant for the $^8$Be anomaly.  In the second and third scenarios, the dark photon is allowed to also couple with fermions in the second family, with equal or opposite signs to the couplings in the first family.  We also consider three schemes for the estimation of experimental reach.

We estimate the projected limits by calculating the numbers of events from both dark photon and SM background.  We perform a scan of dark photon couplings, $(\varepsilon Q_e,\varepsilon Q_p, \varepsilon Q_n)$, and compare the results with the region favored by the $^8$Be anomaly.  
Moreover, we show the angular distributions of final-state $e^+e^-$ and $\mu^+\nu$ systems in their own center mass frames.

In general, Scheme~1 has the best sensitivity to probe the dark photon couplings due to a larger number of events as well as a better energy resolution.  Most of the parameter space preferred by the $^8$Be anomaly can be probed by both $K^+ \to  \mu^+\, \nu_{\mu}\,  e^+ e^-$ and $\pi^+ \to  \mu^+\, \nu_{\mu}\,  e^+ e^-$ decays.  The effects from kaon and pion form factors are found to be negligible in the projected limits for $m_X=16.7$~MeV.  In each specific experimental scheme, the projected limits from the kaon decay are stronger than those from the pion decay.  Except for some special case where $K^+$ and $\mu^+$ have opposite couplings with the dark photon, the signal events and the SM background events have different behaviours in the $\mu^+\nu$ angular distributions.  The $e^+e^-$ angular distribution can directly reveal the vectorial nature of the dark photon interaction.


\section*{Acknowledgments}
C.W.C would like to thank the hospitality of Kavli IPMU during his visit in 2016 summer, where this project was initiated.  This research was supported in part by the Ministry of Science and Technology of Taiwan under Grant No.\ NSC 104-2628-M-002-014-MY4, and in part by World Premier International Research Center Initiative (WPI), MEXT, Japan.

\end{document}